\documentclass[preprint,aps,superscriptaddress,nofootinbib,showkeys,11pt]{revtex4}

\usepackage{graphicx}
\usepackage{amsmath}
\usepackage{amsfonts}
\usepackage{placeins}
\usepackage{mathtools}
\usepackage{todonotes}
\usepackage{physics}
\usepackage{amssymb}
\usepackage{bm}
\usepackage{multirow}
\usepackage{csquotes}
\usepackage{epstopdf}
\usepackage{float}
\usepackage{subfigure}
\usepackage{capt-of}
\allowdisplaybreaks

\usepackage{hyperref}

\hypersetup{
     colorlinks   = true,
     citecolor    = red,
     linkcolor    = blue,
     urlcolor     = blue,
}
\definecolor{deeppink}{rgb}{0.9, 0.17, 0.31}

\def\beq{\begin{equation}}
\def\eeq{\end{equation}}
\def\bea{\begin{align}}
\def\eea{\end{align}}

\def\l{\left}
\def\r{\right}
\def\vk{\vec{k}}
\def\vko{\vec{k}_1}
\def\vks{\vec{k}_2}
\def\vp{\vec{p}}
\def\vx{\vec{x}}

\def\Pt{\mathcal{P}_{\rm T}}
\def\Ps{\mathcal{P}_{\mathcal{S}}}
\def\Pr{\mathcal{P}_{\mathcal{R}}}

\def\wre{w_{\phi }}

\def\nn{\nonumber}

\def\kre{k_{\rm re}}

\def\ere{\eta_{\rm re}}
\def\ke{k_{\rm end}}
\def\He{H_{\rm end}}
\def\HI{H_{\rm end}}

\def\MP{{ M}_{\rm pl}}
\def\mpl{{ M}_{\rm pl}}

\def\kbar{\bar{k}}
\def\Tre{T_{\rm re}}
\def\tre{T_{\rm re}}

\def\Nre{N_{\rm re}}

\def\ogwp{\Omega_{\rm gw}^{\rm pri}}
\def\nw{n_w}

\def\ogws{\Omega_{\rm gw}^{\rm sec}}

\def\mJ_{\mathrm{J}}
\def\nno{\nonumber}
\def\rhor{\rho_{\rm\chi}}
\def\rphi{\rho_{\rm\phi}}
\def\wre{w_{\phi}}
\def\wphi{w_{\rm\phi}}
\def\aend{a_{\rm end}}
\def\hend{H_{\rm end}}
\def\mphi{m^{\rm end}_\phi}

\graphicspath{{./figs/}}


\begin{document}

 \title{
Nonminimal infrared gravitational reheating in light of ACT observation
}

 \author{Ayan Chakraborty}
\email{chakrabo@iitg.ac.in}
\affiliation{Department of Physics, Indian Institute of Technology, Guwahati, 
Assam, India}
\author{Debaprasad Maity}
\email{debu@iitg.ac.in}
\affiliation{Department of Physics, Indian Institute of Technology, Guwahati, 
Assam, India}
\author{Rajesh Mondal}
\email{mrajesh@iitg.ac.in}
\affiliation{Department of Physics, Indian Institute of Technology, Guwahati, 
Assam, India}
 \begin{abstract}
 Inflation is known to produce large infrared scalar fluctuations. Further, if a scalar field $(\chi)$ is non-minimally coupled with gravity through $\xi \chi^2 R$, those infrared modes experience \textit{tachyonic instability} during and after inflation. Those large non-perturbative infrared modes can collectively produce hot Big Bang universe upon their horizon entry during the post-inflationary period. We indeed find that for reheating equation of state (EoS), $w_{\phi} > 1/3$, and coupling strength, $\xi>1/6$, large infrared fluctuations lead to successful reheating. We further analyze perturbative reheating by solving the standard Boltzmann equation
in both Jordan and Einstein frames, and compare the results with the non-perturbative ones. 
Finally, embedding this infrared reheating scenario into the well-known $\alpha-$attractor inflationary model, we examine possible constraints on the model parameters in light of the latest Atacama Cosmology Telescope (ACT), Dark Energy
Spectroscopic Instrument (DESI) results. To arrive at the constraints, we 
take into account the latest bounds on tensor-to-scalar ratio, $r_{0.05}\leq 0.038$, isocurvature power spectrum, $\mathcal{P}_{\mathcal{S}} \lesssim 8.3\times 10^{-11}$, 
and effective number of relativistic degrees of freedom, $\Delta N_{\rm eff} \lesssim 0.17 $. 
Subject to these constraints, we find
successful reheating to occur only for EoS $w_{\phi}\gtrsim 0.6$, which translates to a sub-class of $\alpha-$attractor models being favored and placing them within the 2$\sigma$ region in the $ n_s-r$ plane of the latest ACT, DESI data. 
In this range of EoS, we find that the coupling strength should lie within $2.11\lesssim\xi\lesssim 2.95$ for $w_{\phi}=0.6$, and for $w_{\phi}>0.6$, the allowed range becomes even tighter. Finally, we compute secondary gravitational wave signals induced by the scalar infrared modes, which are found to be strong enough to be detected by future GW observatories, namely BBO, DECIGO, LISA, and ET.    
\end{abstract}
 
\maketitle

\section{Introduction}
Inflation is considered to be the most successful paradigm in explaining large-scale observations of our universe. Over the years large number of models have been proposed to explain such phenomena. Latest precision observation by ACT, DESI \cite{ACT:2025fju, ACT:2025tim}, combined with \textit{Planck}, however, seemingly make a large number of well-motivated inflationary models disfavored. However, inference on any inflationary model based on observation must include its post-inflationary dynamics. It is important to note that the predictions of any inflationary model, and their relation to the CMB anisotropies, are intimately tied to the post-inflationary reheating phase. This intriguing connection was first explored by Kamionkowski et al. in \cite{PhysRevLett.113.041302}. Consequently, CMB observations, together with the underlying reheating mechanism, play a crucial role in constraining any given inflationary model.
In this light, any new reheating scenario—such as the one we present in this paper—should be examined through the lens of CMB observations in order to obtain the correct constraints on inflationary models. Post-inflationary reheating is considered to be an integral part of the early universe dynamics \cite{Kolb:1990vq,Shtanov:1994ce,Kaiser:1995fb,Kofman:1994rk, Kofman:1997yn,Allahverdi:2010xz,Sato:2015dga, Bassett_2006, Lozanov:2019jxc} connecting \textit{inflation} \cite{Guth:1980zm, Senatore:2016aui,Linde:1981mu,Albrecht:1982wi,Starobinsky:1980te, Lemoine:2008zz, martin2003inflation, Martin, sriramkumar2009introduction,Linde:2014nna,  Piattella:2018hvi,Baumann_2009, Baumann:2018muz} and the \textit{hot, thermal radiation-dominated phase}. 
In this paper, we propose a new gravitational reheating mechanism, discuss its indirect impact, and revisit the possible constraint on the parameter space of $\alpha-$attractor model in light of the latest ACT, DESI observations. Such observation leads to a surge of activity on the inflationary model building and revisiting the existing models (see \cite{Kallosh:2025rni, Aoki:2025wld, zharov2025reheatingactsstarobinskyhiggs, liu2025reconcilinghiggsinflationact, Yogesh:2025wak, haque2025actdr6insightsinflationary, Haque:2025uga, Mondal:2025kur, Maity:2025czp, Drees:2025ngb, He:2025bli, yin2025higgslikeinflationactivatedmass, gialamas2025keepingrelationstarobinskymodel, Dioguardi:2025mpp, Dioguardi:2025vci, Gialamas:2025kef}).

In the conventional reheating scenarios, inflaton is modeled to decay into the radiation field through its direct coupling. The dynamics can be dominated by perturbative \cite{Kolb:2003ke,Chung:1998rq,Giudice:2000ex,Mukaida:2012bz,Maity:2018dgy,Drewes:2017fmn,Garcia:2020eof,Drewes:2013iaa,Drewes:2015coa,Garcia:2020wiy,Adshead:2019uwj,Haque:2023yra,Haque:2020zco,Adshead:2016xxj}, non-perturbative \cite{Kofman:1994rk, Kofman:1997yn,Greene:1997fu, Amin:2014eta,Maity:2018qhi, Chakraborty:2023lpr, Tsujikawa:1999jh,Tsujikawa:1999iv, Dufaux:2006ee, Abolhasani:2009nb, Greene:1998nh, Greene:2000ew, Peloso:2000hy}, or both depending on the strength of the coupling between the inflaton and radiation field. However, it is important to realize that all those processes are causal and hence deal with the modes that live inside the Hubble horizon. In terms of this conventional decay process, it is very difficult to think that the modes that are super-horizon can be produced, and will have any impact on the reheating process.
Such a question has already been raised and discussed earlier in the literature \cite{Ford:2021syk,Chakraborty:2024rgl}. It is well known that quantum mechanically, purely de Sitter spacetime is unstable under super-horizon perturbations, which typically manifests itself in terms of infrared divergence \cite{Cespedes:2023aal}. However, such divergences are likely to be absent if the de Sitter phase survives for a short period, such as the inflationary phase in the early universe. During this early universe inflationary phase, large infrared fluctuations of any light fields are indeed produced and can have a significant impact on the after-inflation dynamics, such as reheating \cite{Opferkuch:2019zbd,Bettoni:2021zhq,Laverda:2023uqv, Figueroa:2024yja,Markkanen:2015xuw, Markkanen:2017edu}. In most of the previous reheating analyses, such super-horizon modes are ignored due to their very acausal nature. Production of super-horizon perturbations, therefore, is a unique feature of inflation which is indeed observed in Cosmic Microwave Background(CMB) temperature anisotropy. Those are identified with the massless inflaton fluctuation. However, at the inflation scale of order $10^{15}$ GeV, all the Standard Model (SM) fields are massless and can be generated amply at super-horizon scales during inflation. Such inflationary super-horizon production, therefore, becomes sub-horizon in the post-inflationary period and can contribute to subsequent dynamics of the universe.

In the reference (see \cite{Ford:2021syk}), such a contribution has been taken into account for the first time and demonstrates the possibility of successful reheating, particularly for a stiff reheating equation of state. This has been further studied in the context of reheating in the kination regime \cite{Dimopoulos_2018, Lankinen_2017,  Lankinen:2019ifa}. In this paper, we generalize those in the context of non-minimal gravity.  
For the minimally coupled theory, after their horizon exit during inflation, the super-horizon modes of massless fields remain constant until their reentry during the standard Big Bang evolution. Therefore, such a scenario generally predicts low reheating temperature \cite{Haque:2022kez, Clery:2021bwz, Chakraborty:2025zgx}. 
However, it has been observed that for a non-minimally coupled scalar field namely $\xi \chi^2 R$, the super-horizon modes can grow due to \textit{tachyonic instability} \cite{Chakraborty:2024rgl,Figueroa:2021iwm,Opferkuch:2019zbd,Bettoni:2021zhq,Laverda:2023uqv, Figueroa:2024yja,Laverda:2024qjt,Markkanen:2015xuw, Markkanen:2017edu, Fairbairn:2018bsw, Dimopoulos_2018}. In this paper, we shall demonstrate that such modes reentering the horizon after the conclusion of inflation can successfully reheat the universe without any further coupling parameter in the inflaton sector and predict a high reheating temperature. Since the reheating is solely aided by the infrared modes and produced by non-minimal gravitational interaction, we shall call it \enquote{Non-minimal Infrared Gravitational Reheating}. 

Reheating governed by purely gravitational interaction has recently been discussed in the literature  \cite{Haque:2022kez,Clery:2022wib,Dimopoulos_2018,Nakama:2018gll}. In all these studies, the focus has been on sub-Hubble modes during the reheating phase. The dynamics of these modes are well described by the standard Boltzmann equation, which assumes inflaton decay into massless radiation. This decay process is mediated by gravitons through minimal gravitational interactions of the form \((1/M_{\rm pl}) h_{\mu\nu} T^{\mu\nu}_{\chi}\). Such scenarios have gained significant interest due to their universal nature and model-independent predictions, particularly in determining the reheating temperature and dark matter mass. The predictions of gravitational reheating scenarios depend entirely on inflationary parameters, particularly the inflaton equation of state (EoS) \( w_{\phi} \). For instance, if the inflaton potential follows a standard power-law form near its minimum, \( V(\phi) \sim \phi^{2n} \), the effective inflaton EoS, given by \( w_{\phi} = (n-1)/(n+1) \), governs the entire reheating dynamics. However, this compelling scenario is ruled out for any value of \( 0\leq w_\phi\leq 1 \) due to constraints from Big Bang nucleosynthesis (BBN)\footnote{Recently proposed gravitational neutrino reheating \cite{Haque:2023zhb,Haque:2024zdq} scenarios alleviate such constraints, making the framework viable for any inflaton EoS \( \wre > 1/3 \).}. Specifically, it is inconsistent with the observed effective number of relativistic degrees of freedom, \( \Delta N_{\rm eff} \sim 0.284 \) \cite{Planck:2018jri, Planck:2018vyg} latest observation, and the lower bound on the reheating temperature, \( T_{\rm re}^{\rm min} \sim T_{\rm BBN} \simeq 4 \) MeV. \cite{Kawasaki:2000en, Sarkar_1996, Hannestad:2004px}. Subsequently, the model with non-minimal coupling where the radiation field is gravitationally coupled is studied \cite{Clery:2022wib, Barman:2022qgt} to evade such a problem. It is indeed demonstrated that for a sufficiently large value of $\xi$ the universe can be reheated with $w_\phi > 1/3$ in consistent with Big Bang nucleosynthesis(BBN) observation. One should remember that in all the aforesaid studies, the standard Boltzmann framework has been adopted, which deals with only the sub-Hubble modes of the fluctuation.    

In this paper, we shall compare two different production mechanisms stated above. We analyze the contribution of both super- and sub-Hubble modes, which are produced during and after inflation in the process of reheating. Our analysis reveals that since reheating is a gradual process, a large number of super-horizon modes will enter the horizon during this process, resulting in a non-negligible contribution compared to the modes causally produced via Boltzmann dynamics from the inflaton decay. Indeed, in some regions of parameter space, we show that the energy density associated with the inflationary super-horizon modes of radiation field non-minimally coupled with gravity $\xi\chi^2R$ can supersede the contribution from their causal counterpart produced solely from the inflaton decay.

The order of construction of this paper is as follows: In Section \ref{sec2}, we first introduce the non-perturbative framework of gravitational particle production in the presence of non-minimal gravity coupling. In Section \ref{sec3}, we study the non-perturbative dynamics of the \textit{infrared gravitational reheating}. In Section \ref{sec4}, we do a comparative study between the non-minimal coupling-induced perturbative gravitational reheating and the non-perturbative infrared reheating. 
In Section \ref{secgw}, we show that based on observational bounds on the tensor-to-scalar ratio and the isocurvature perturbation amplitude, there exists an upper limit on the coupling strength ($\xi_{\rm max}$). In Section \ref{secgwsign}, considering the induced gravitational wave we further obtain a lower limit on the coupling strength($\xi_{\rm min}$) based on the $\Delta N_{\rm eff}$ bound for the primary gravitational wave(PGW). Finally we identify the region of $\xi$ vs $\alpha$ and $\xi$ vs $T_{\rm re}$ parameter spaces which are fully consistent with all the observational bounds and latest ACT results.  
Finally, Section \ref{secconclusion} concludes the paper by giving a brief outline of the main outcomes of this study. In Appendix \ref{appenproductionrate}, we detail the computation of the minimal and non-minimal perturbative production rates of the massless scalar fluctuation. In Appendix \ref{appenA}, we compute the expression of the isocurvature perturbation amplitude for massless fluctuations. In Appendix \ref{appenJorEin}, 
we do a comparative analysis between the Jordan and Einstein frame $\Tre$ vs $\xi$ predictions for different reheating EoS.  

\maketitle

\section{Spectrum of Gravitationally Produced Massless Particles}\label{sec2}

Assuming a massless scalar field $\chi$ as radiation non-minimally coupled to gravity, we shall begin with the following inflaton($\phi$)-radiation system 
\begin{equation}\label{lagrangian1}
    \mathcal{L}_{[\phi,\chi]}= -\sqrt{-g}\bigg(\frac{1}{2}\partial_{\mu}\phi \partial^{\mu}\phi+V(\phi)+\frac{1}{2}\partial_{\mu}\chi \partial^{\mu}\chi+\frac{1}{2}
    \xi R\chi^2 \bigg) .
\end{equation}
Where the  Friedmann-Lemaître-Robertsom-Walker (FLRW) metric is expressed as $ds^2=a^2(\eta)\big(-d\eta^2+d\vec{x}^2\big)$ with the scale factor $a$ and $\sqrt{-g}=a^4(\eta)$ 
. $V(\phi)$ is the inflaton potential, 
\enquote{$\xi$} is the dimensionless non-minimal coupling of $\chi$ field with gravity  Ricci scalar \enquote{$R$} generates a time-dependent effective mass for the $\chi$ field as, $m_{\text{eff}}^2(\eta)=\big(m_{\chi}^2+\xi R(\eta)\big)$.


Expressing the scalar field \enquote{$\chi$} in terms of Fourier modes,
\begin{equation}\label{fourier}
     \chi(\eta,\vec{x})= \int\frac{d^3\vec{k}}{(2\pi)^3} ~\chi_{\vec{k}}(\eta) e^{i\vec{k} \vec{x}},
 \end{equation}
 and subject to the Lagrangian (\ref{lagrangian1}) we reach the following dynamical equation of mode function ($\chi_{\vec{k}})$ as,
   \begin{equation}\label{dynamical1}       \chi_{\vec{k}}^{\prime\prime}+2\mathcal{H}\chi_{\vec{k}}^{\prime}+\Big(k^2+a^2(\eta)\xi R\Big)\chi_{\vec{k}}=0 .
  \end{equation}
  In the dynamical Eq. (\ref{dynamical1}), there is a damping term, \enquote{$2\mathcal{H} \chi_{\vec{k}}^{\prime}$} with conformal Hubble scale $\mathcal{H}$, which is non-zero in expanding background. Defining a new rescaled field $X_{\vec{k}} = a(\eta)\chi_{\vec{k}}(\eta)$, we can transform the above equation into the following simple form of harmonic oscillator with time-dependent frequency,
   \begin{align}\label{dynamical2}       
   X_{\vec{k}}^{\prime\prime}+\omega_k^2(\eta) X_{\vec{k}}=0 .
  \end{align}
The time-dependent frequency \enquote{$\omega_k$} is expressed as 
 $\omega_k^2(\eta)=k^2-{a^2 R}(1/6-\xi)$. 
Note that in the conformal limit $\xi = 1/6$, the field effectively becomes massless. To solve the Eq. (\ref{dynamical2}) we choose the positive frequency Bunch-Davies vacuum when the modes live deep inside the horizon,
\begin{equation}\label{Bunchdavies}
   X_{k}(\eta_0)=\frac{1}{\sqrt{2\omega_k(\eta_0)}}e^{-i\omega_k\eta_0}, \quad X_{k}^{\prime}(\eta_0)=-i\sqrt{\frac{\omega_k(\eta_0)}{2}}e^{-i\omega_k\eta_0}.
   \end{equation}
Where $\eta_0\rightarrow -\infty$ is the initial time(beginning of inflation) when positive-frequency Bunch-Davies vacuum solution is satisfied. The particle occupation number density power spectrum for the scalar field is usually expressed as \cite{Kofman:1997yn}, 
 \begin{equation}\label{fnoden}
     n_k=\frac{1}{2\omega_k}|\omega_kX_k-iX_k^{\prime}|^2 .
 \end{equation}

Integrating Eq.(\ref{fnoden}) over all the momentum modes we get the total number and UV convergent energy density as \cite{PhysRevD.9.341,PhysRevD.36.2963,Kolb:2023ydq}
\begin{align}\label{fnoEden}
 n_{\chi}=\frac{1}{(2\pi)^3a^3} \int d^3k n_k, ~~;~~
 \rho_{\chi}=\frac{1}{(2\pi)^3a^4} \int d^3k\omega_k n_k .
\end{align}
Utilizing this non-perturbative formalism outlied above, we analyze in detail the tachyonic growth of a non-minimally coupled scalar field and its impact on the reheating of the universe. 

The evolution of scale factor during inflation and any general reheating EoS can be represented as a function of conformal time as 
\begin{align}\label{scalefactor}
     a(\eta)=
     \begin{cases}
        -\frac{1}{H_{\text{end}}\eta}\quad \quad \quad -\infty<\eta\leq \eta_{\text{end}} & \\
a_{\text{end}}\Big(\frac{1+3\wre}{2|\eta_{\rm end}|}\Big)^{\frac{2}{1+3\wre}}\bigg(\eta-\eta_{\text{end}}+\frac{2 |\eta_{\rm end}|}{1+3\wre}\bigg)^{\frac{2 }{1+3\wre}}  \eta\geq\eta_{\text{end}} .&
\end{cases}
\end{align}
Considering pure de Sitter inflation, we assume $H_{\text{ds}}=H_{\text{end}}$. 
It is straightforward to check that during the transition from inflation to reheating, the scale factor and its first derivative change continuously at the junction point, that is at the end of inflation, $\eta=\eta_{\text{end}}=-(1/a_{\rm end}H_{\rm end})$. Here $a_{\rm end}$ is the scale factor at $\eta=\eta_{\rm end}$ and $\wre$ is the background inflaton EoS during reheating.

For our later purpose, we express the Hubble scale in term of inflaton equation of state as,
\begin{equation}\label{hubble}
 \mathcal{H}(\eta\geq \eta_{\rm end})=\frac{a^{\prime}(\eta)}{a(\eta)}=\frac{2 (a_{\rm end}H_{\rm end})}{(1+3\wre)}\bigg((\eta a_{\rm end} H_{\rm end})+\frac{3(1+\wre)}{(1+3\wre)}\bigg)^{-1} . \end{equation}
Where $H_{\rm end}$ is the Hubble scale at the inflation end.

\begin{figure}[t]
     \begin{center}
\includegraphics[scale=0.35]{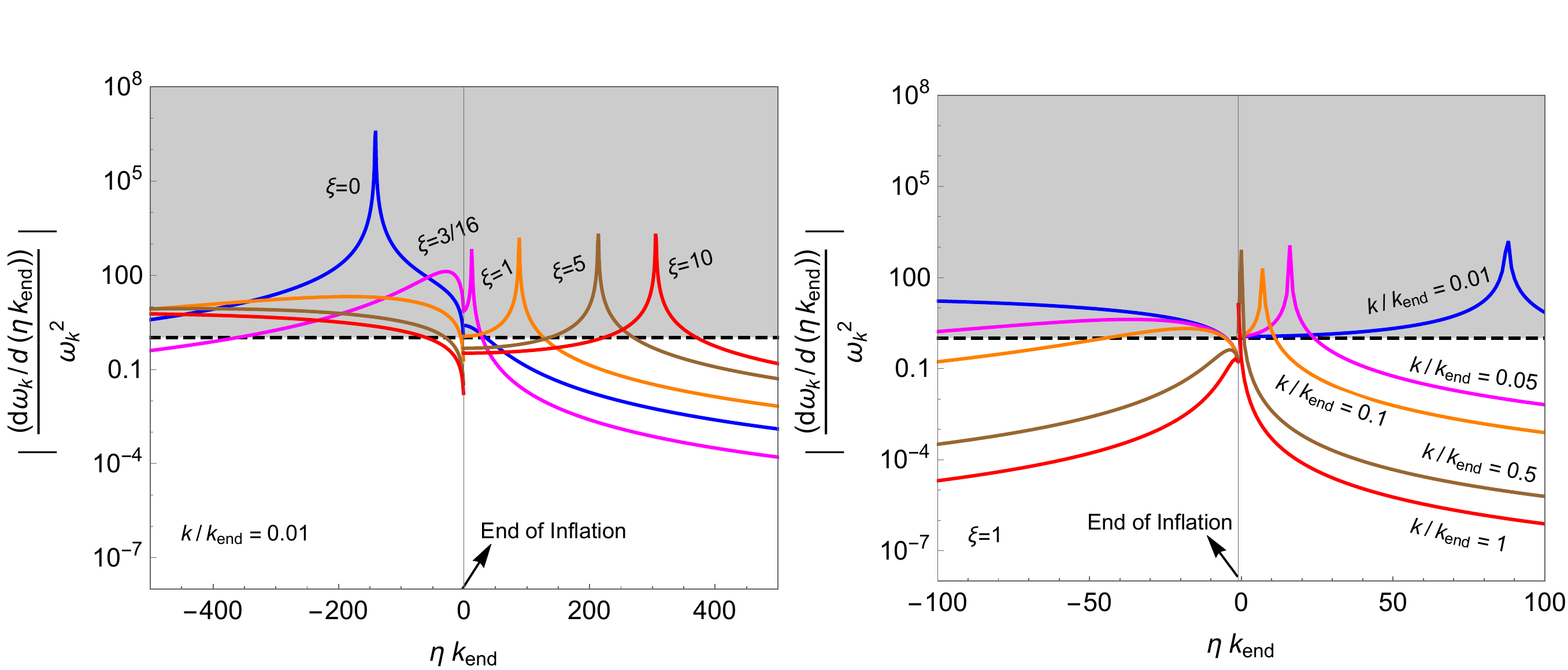}
\caption{\textit{Figure represents the measure of adiabaticity violation in terms of parameter $|(d\omega_k/d(\eta k_{\rm end})/\omega_k^2|$ with $\eta k_{\rm end}$ for different coupling strengths $\xi$ (left panel) and different scales $k/k_{\rm end}$ (right panel) for a specific EoS $\wre=1/2$. 
In both panels, the black dashed line indicates adiabaticity parameter, $|(d\omega_k/d(\eta k_{\rm end})/\omega_k^2|=1$. Any value of $|(d\omega_k/d(\eta k_{\rm end})/\omega_k^2|>1$ depicted by the gray shaded region indicates the violation of adiabaticity. In the left panel of this figure, for the given scale $k/k_{\rm end}=0.01$, with the increase of $\xi$ values, the peak of the adiabaticity parameter gradually shifts from the inflationary to the post-inflationary phase. This indicates that the super-horizon modes can still grow during reheating for higher coupling $\xi$. For $\xi=0$, the instability effect is only present in the inflationary phase. In the right panel, it shows that for a given non-minimal coupling $\xi$ with the increase of $k/k_{\rm end}$ (small scale), the modes tend to remain adiabatic through the evolution. 
}}
\label{adiabaticityviolationfig}
\end{center}
\end{figure}

Inflationary spacetime has an appealing property that it behaves almost adiabatically in the asymptotic past and future. In these asymptotic limits, spacetime evolves very slowly, thereby causing the fluctuations on top of this background to be adiabatic in the asymptotic past and future. In particular, one defines a dimensionless parameter $|\omega^{\prime}_k/\omega_k^2|$ to study the departure from the adiabatic limit. It can be shown explicitly as $\eta\rightarrow \pm\infty$, the ratio approaches $|\omega^{\prime}_k/\omega_k^2|\rightarrow 0$ sufficiently fast. In the process of transition of the universe from early de Sitter to some post-inflationary phase, this adiabaticity condition gets violated ($|\omega^{\prime}_k/\omega_k^2|>>1$) at some intermediate point (See Fig.\ref{adiabaticityviolationfig}), and causes particle production associated with long-wavelength modes. However, small scales $k > a_{\rm end} H_{\rm end}$ residing inside the horizon generally remain adiabatic without any parametric growth. Therefore, after the inflation super-Hubble modes are expected to contribute most to the total energy density after horizon reentry during reheating.
Let us suppose $X_k^{(\rm inf)}(\eta)$ is the adiabatic vacuum solution of (\ref{dynamical2}) during the de Sitter phase in the time interval $-\infty<\eta\leq\eta_{\text{end}}$ and $X_k^{(\rm reh)}(\eta)$ is the adiabatic vacuum solution during reheating phase for $\eta\geq\eta_{\text{end}}$. Making these solutions and their first derivatives continuous at the junction $\eta=\eta_{\text{end}}$, 
we compute the Bogoliubov coefficients $\alpha_k, \beta_k$  as follows:  \cite{deGarciaMaia:1993ck, Kolb:2023ydq}
\begin{align}\label{bogo}
  &\alpha_k= i\left({X_k^{(\rm inf)}}'(\eta_{\rm end}){X_k^{(\rm reh)}}^{*}(\eta_{\rm end})-{X_k^{(\rm inf)}}(\eta_{\rm end}){X_k^{(\rm reh)}}^{*\prime}(\eta_{\rm end})\right)\nonumber\\
  & \beta_k=-i\left({X_k^{(\rm inf)}}'(\eta_{\rm end}) X_k^{(\rm reh)}(\eta_{\rm end}) - {X_k^{(\rm reh)}}'(\eta_{\rm end}) X_k^{(\rm inf)}(\eta_{\rm end})\right)
  \end{align}
where ($'$) denotes the derivative with respect to conformal time and both the vacuum solutions in the above Eq.(\ref{bogo}) satisfy the Wronskian condition $\big({X_k^{(\rm inf(reh))}}{X_k^{(\rm inf( reh))}}^{*\prime}-{X_k^{(\rm inf(reh))}}'{X_k^{(\rm inf(reh))}}^{*}\big)=i$ at any time $\eta$. Any general field solution during reheating can thus be expressed as $X_k(\eta)=\l(\alpha_k X_k^{(\rm reh)}+\beta_k X_k^{*(\rm reh )}\r).$
The appearance of non-zero $\beta_k$, caused by the breakdown of the adiabaticity condition, results in the mixing of positive and negative-frequency modes in the general field solution of the post-inflationary phase. 

Using the scale factor (\ref{scalefactor}) in the  equation (\ref{dynamical2}), we obtain the following two different adiabatic solutions in two different phases \cite{Chakraborty:2024rgl}, 
\begin{equation}\label{Xinf3}
    X_k^{(\rm inf)}=\frac{\sqrt{-\pi \eta}}{2}e^{i(\pi/4+\pi\nu_1/2)}H^{(1)}_{\nu_1}(-k\eta)~~~;~~~ X_k^{(\rm reh)}(\eta)=\sqrt{\frac{\bar{\eta}}{\pi}}\text{exp}\bigg[\frac{3ik \mu}{k_{\rm end}}+\frac{i\pi}{4}\bigg]K_{\nu_2}(i k \bar{\eta}).
\end{equation}
Where $H^{(1)}_{\nu_1}$ is the Hankel function of first kind of order $\nu_1$,
and $K_{\nu_2}$ is the modified Bessel function of second kind order $\nu_2$.
$k_{\rm end}=a_{\rm end}H_{\rm end}$ is the scale that leaves the horizon at the end of inflation. We use the symbol $\bar{\eta} = (\eta + {3\mu}/a_{\rm end}H_{\text{end}})$.  Expressions of all the sybmols are 
\begin{align}\label{symbol}
   &\mu=\frac{(1+\wre)}{(1+3\wre)}, \quad \nu_1=\frac{3}{2}\sqrt{1-\frac{16}{3}\xi}, \quad 
\nu_2=\frac {3(1-\wre)}{2(1+3\wre)}{\sqrt{
1-\frac{16}{3}\xi\frac{(1-3\wre)}{(1-\wre)^2}}
}  .
\end{align}
Depending upon the value of the non-minimal coupling constant $\xi$, the order of the inflationary vacuum solution $\nu_1$ becomes positive, taking values from $3/2$ to $0$ for $0\leq\xi<3/16$, where zero for $\xi=3/16$. Obviously $\nu_1$ becomes imaginary for $\xi>3/16$. Interestingly, the index of post-inflationary vacuum solution $\nu_2$ also becomes imaginary in the range $\xi>3/16$ for $0\leq\wre<1/3$ and becomes real positive for $1/3\leq\wre\leq 1$, and it ranges within $1/2 \leq \nu_2 \leq \sqrt {3\xi/2}$. This varying nature of the indices $\nu_1, \nu_2$ depending on non-minimal coupling ($\xi$) and post-inflationary EoS $\wre$ causes non-trivial modification to the Bogoliubov coefficients $\alpha_k,\beta_k$ as well as the nature of the post-inflationary field solution. Now, we shall compute the number-density spectra of produced massless particles for general post-inflationary EoS in three specified ranges of the non-minimal coupling strength $\xi$ stated above. With these amplified infrared fluctuations, we now study the reheating dynamics.

\subsection{Comoving number density spectrum ($|\beta_k|^2$) }
\begin{figure}[t]
     \begin{center}
\includegraphics[scale=0.25]{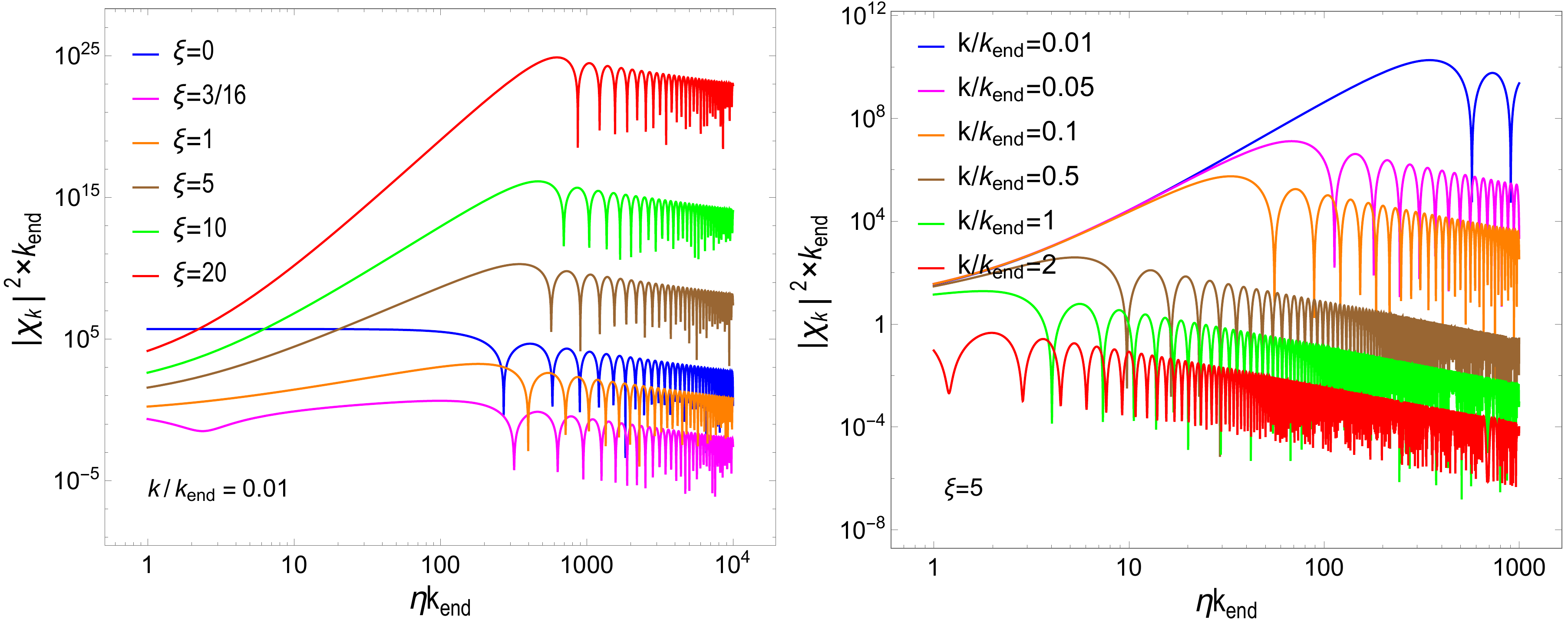}
\caption{\textit{Figure represents the variation of dimensionless field variable $|\chi_k|^2 \times\ke$ with dimensionless time $\eta k_{\rm end}$ in the post-inflationary phase. The left panel of the figure shows that with the increase of $\xi$, the post-inflationary instability effect becomes vital for a particular super-Hubble mode at the inflation end. The super-horizon growth will sustain until a particular mode enters the horizon and after horizon entry, the field mode will be oscillatory. In the right panel of the figure, it is observed that the longer the wavelength, the stronger the instability growth for a particular coupling strength, and this is true for any non-minimal coupling strength $\xi$. }}
\label{fieldvstimefig}
\end{center}
\end{figure}
From the Fig.(\ref{adiabaticityviolationfig}), one can read the fact that due to non-minimal coupling, a tachyonic instability is indeed developed both during and after inflation. When the modes cross the horizon during inflation, namely the super-horizon modes, the instability effect ($\omega_k^2<0$) is seen to be appreciable only for $0\leq\xi<1/6$.
Interestingly, as $\xi$ exceeds the conformal limit $\xi=1/6$, this instability gradually becomes insignificant during inflation, and becomes significant in the post-inflationary era, particularly for $\wre>1/3$ 
(See Fig.(\ref{fieldvstimefig}) and the references \cite{Chakraborty:2024rgl, Fairbairn:2018bsw} for detailed discussion). The growth of the field amplitude is also connected to the enhancement of the $|\beta_k|^2$ spectrum through the relation (\ref{fnoden}).
We now calculate the number density of those infrared modes, which is defined once those modes enter the horizon during the post-inflationary period. The spectral behaviour of $|\beta_k|^2$ is required to calculate the energy density, which we shall compute in the next section while studying the \textit{infrared gravitational reheating} dynamics.
As is obvious from Fig.(\ref{fieldvstimefig}) that the enhancement is not effective for shorter wavelengths ($k \gtrsim k_{\rm end}$) modes. Therefore, we calculate the number spectrum for modes lying in the range $k_{\rm re}<k<k_{\rm end}$ where $k_{\rm re}=a_{\rm re}H_{\rm re}$ is the mode which enters the Hubble horizon at the end of reheating, and $a_{\rm re},\,H_{\rm re}$ are the scale factor and Hubble scale at the end of reheating. For different ranges of $\xi$, the spectral behaviors of the comoving number densities are as follows:

\noindent
\underline{For $0\leq\wre<1/3$ :}
\begin{equation}\label{no.den1}
   |\beta_k|^2\propto
   \begin{cases}
       (k/k_{\rm end})^{-2(\nu_1+\nu_2)} \quad &\text{for} \quad 0\leq\xi<3/16 \\
       (k/k_{\rm end})^{-2\nu_2} \quad &\text{for} \quad \xi=3/16 \\
        \end{cases} 
\end{equation}
\underline{For $1/3\leq\wre\leq1$ :}
\begin{equation}\label{no.den2}
   |\beta_k|^2\propto
   \begin{cases}
       (k/k_{\rm end})^{-2(\nu_1+\nu_2)} \quad &\text{for} \quad 0\leq\xi<3/16 \\
       (k/k_{\rm end})^{-2\nu_2} \quad &\text{for} \quad \xi=3/16 \\
       (k/k_{\rm end})^{-2\nu_2} \quad &\text{for} \quad \xi>3/16 
        \end{cases} 
\end{equation}
Depending upon different post-inflationary EoS, the number density spectrum follows different power-law behavior \cite{Chakraborty:2024rgl}. It is important to note that for $0\leq\wre<1/3$, in the range $\xi>3/16$, we cannot obtain a nice power-law behavior of the spectrum like the other cases due to the imaginary nature of both the indices $\nu_1$ and $\nu_2$. For this particular case, the spectral behavior is found in \cite{Chakraborty:2024rgl}.\\ 
Based on these comoving number density spectra, we now analyze the possibility of reheating.

\section{Infrared Gravitational Reheating: Defining Reheating parameters ($N_{\rm re}$, $T_{\rm re}$)}\label{sec3}
In the previous section, we have discussed the non-trivial effect of non-minimal coupling on the dynamics of long-wavelength modes of massless scalar fluctuations during inflation and post-inflationary phases \cite{Chakraborty:2024rgl}. 
Such super-horizon growth beyond a certain threshold of $\xi$ essentially directs us to investigate the possibility of successful reheating without invoking any further new physics in the inflaton sector. The non-minimal coupling between gravity and radiation provides us an extra gravitational parameter $\xi$, which can be tuned to obtain radiation energy density that can surpass the contribution from the sub-Hubble modes, and lead to successful reheating. 
Therefore, controlling $\xi$ can give rise to large reheating temperature as compared to the pure gravitational reheating scenario \cite{Haque:2022kez, Haque:2023yra, Chakraborty:2023ocr}. In the present section, exploiting all the number density spectra derived in the last Section \ref{sec2} for $\wre>1/3$ (see Eq.(\ref{no.den2})), we shall thoroughly investigate the reheating dynamics by these infrared scalar fluctuations, and also derive the important expressions of reheating parameters ($N_{\rm re}, T_{\rm re}$) in different ranges of coupling parameter $\xi$.

For minimally coupled inflaton without any additional interaction, inflaton energy density scales as $\rho_{\phi}\propto a^{-3(1+w_{\phi})}$ whereas produced massless fluctuations being radiation scales as $\rho_{\chi}\propto a^{-4}$. To successfully reheat the universe, $\rho_{\phi}$ and $\rho_{\chi}$ must be equal at a point where reheating is assumed to end and standard radiation domination starts.
Note from the Fig.\ref{adiabaticityviolationfig} that adiabaticity violation occurs for a brief period of time, and hence like conventional perturbative reheating, production of infrared radiation is not a gradual process. This essentially suggests that such infrared radiation can dominate over the background inflaton only for $\wre >1/3$.      
Therefore, in the subsequent part of the discussion on energy density, we shall only focus on the spectrum for $\wre>1/3$. 

\subsection{$\alpha-$attractor E-model}
To proceed we shall consider the well known $\alpha$-attractor E-model with the potential \cite{Martin:2013tda, Martin:2013nzq, Cook:2015vqa, Drewes:2017fmn},
\begin{equation}\label{alphapotential}
V(\phi)=\Lambda^4\Big(1-e^{-\sqrt{\frac{2}{3\alpha}}.\frac{\phi}{M_{\rm pl}}}\Big)^{2n} .\end{equation}
where $M_{\rm pl}={1}/{\sqrt{8\pi G}}\approx 2.435\times 10^{18}$ GeV is the reduced \textit{Planck} mass. By varying the exponent \enquote{$n$} we achieve different power law forms of the potential namely, quadratic model(for $n=1$), quartic model (for $n=2$), and so on. The post-inflationary background average EoS is also expressed in terms of the exponent
\begin{eqnarray}\label{eq:wphi}
    \wre=\frac{(n-1)}{(n+1)} \implies n = \frac{(1+\wre)}{(1-\wre)}.
\end{eqnarray}
It is worth mentioning that for sufficiently large values of the inflaton field, the above potential maintains a plateau region necessary for the inflationary phase to occur. The amplitude of the potential \enquote{$\Lambda$}, which measures the energy content in the inflaton during inflation, is constrained by the CMB measurement, and is related to the scalar spectral index $n_s$, the amplitude of the inflaton fluctuation measured as CMB normalization $A_s = 2.1\times 10^{-9}$ and tensor to scalar ratio $r$. The model is favored by the latest \textit{Planck}, ACT, DESI, and
 BICEP/Keck combined (P+ACT+LB+BK18) observational data sets (see the references \cite{ACT:2025tim, ACT:2025fju}),
 where $n_s=0.9743\pm 0.0034$ at $68\%$ C.L. and $95\%$ C.L. upper limit on tensor-to-scalar ratio $r_{0.05}$ is obtained as $r_{0.05}<0.038$.
The tensor-to-scalar ratio $r$ can be analytically expressed as,
\begin{equation}\label{rprimary}
 r= \frac{192\alpha n^2(1-n_s)^2}{\Big[4n+\sqrt{16n^2+24\alpha n(1-n_s)(1+n)}\Big]^2}
 \end{equation}
derived in \cite{Drewes:2017fmn}. 
Another parameter $(\alpha)$ determines the shape of the potential. The energy scale of inflation related to the parameter $\Lambda$ can be analytically expressed in terms of CMB parameters as \cite{Drewes:2017fmn}.
\begin{equation}\label{Lambda}
      \Lambda= M_{\rm pl}\Big(\frac{3\pi^2rA_s}{2}\Big)^{\frac{1}{4}}\Bigg[\frac{2n(1+2n)+\sqrt{4n^2+6\alpha(1+n)(1-n_s)}}{4n(1+n)}\Bigg]^{\frac{n}{2}}
   \end{equation}
During inflation, the inflaton satisfies usual slow-roll conditions. The usual condition for the end of inflation is set by one of the slow roll parameters $\epsilon\propto \Big({V'}/{V}\Big)^2$ to be unity. In the context of the potential (\ref{alphapotential}), we can derive an expression of the field value for general $n$ at which inflation ends is
\begin{equation}\label{phiend}
      \phi_{\text{end}}=\sqrt{\frac{3\alpha}{2}}M_{\rm pl} \ln \Big(\frac{2n}{\sqrt{3\alpha}}+1\Big),
\end{equation}
 Using this field amplitude in (\ref{alphapotential}), we have the potential at the inflation end as
\begin{equation}\label{Vend}
    V_{\text{end}}=\Lambda^4\bigg(\frac{2n}{2n+\sqrt{3\alpha}}\bigg)^{2n} 
 \end{equation}
Using (\ref{Vend}), the Hubble scale at the end of inflation is defined as
\begin{equation}\label{Hend}
  H_{\rm end}=\sqrt{\frac{V_{\rm end}}{2M_{pl}^2}}  
\end{equation}
All these inflationary parameters are essential ingredients in the study of reheating dynamics later.
\subsection{Reheating parameters $N_{\rm re}, T_{\rm re}$ for $0\leq\xi<3/16$}
In this parameter range of interest we have already seen that the number spectrum behaves as $|\beta_k|^2\propto (k/k_{\rm end})^{-2(\nu_1+\nu_2)}$ (see Eq.\ref{no.den2}). Utilizing the spectrum, the total comoving energy density is computed to be
\begin{align}\label{comovingenergy1}
&\rho^{\rm com}_{\chi}=\rho_{\chi}\left(\frac{a}{a_{\rm end}}\right)^4=\frac{1}{2\pi^2}\int_{\kre}^{\ke} k^4 |\beta_k|^2d({\rm ln}(k))
\approx \frac{\mathcal{A}_1 \He^4}{4\pi^2\left(2-(\nu_1+\nu_2)\right)} \\
& \mbox{Where},~~~\mathcal{A}_1=\left(\frac{\Gamma(\nu_1)\Gamma(\nu_2)2^{\nu_1}}{8\pi}\left(\frac{2}{3\mu-1}\right)^{\nu_2}\left(\frac{3\mu(1-2\nu_1)+2(\nu_1-\nu_2)}{\sqrt{(3\mu-1)}}\right)\right)^2
\end{align}
For $\wre>1/3$, in this specified range of $\xi$, we always have $(4-2(\nu_1+\nu_2))>0$. Therefore, the maximum contribution to energy is coming from largest mode $\ke$ corresponding to the end of inflation. This property of the blue-tilted spectrum is used to reach the final expression of $\rho^{\rm com}_{\chi}$ in Eq.(\ref{comovingenergy1}). \textit{The highest accessible scale in the finite time scale of reheating is $\kre$ and the modes in the range $\kre\leq k\leq\ke$ are well inside the horizon at the end of reheating, 
and hence contribute to the total energy density of produced particles. This is why $\kre$ is considered to be the IR limit in the computation of the above integration.}

Reheating will be concluded when background energy density will be equal to total radiation energy density $\rho_{\chi}$, the reheating parameters, namely the reheating e-folding number $N_{\rm re}=\text{ln}\left(a_{\rm re}/a_{\rm end}\right)$, is calculated as
\begin{align}\label{Nre1}
    \rho_{\rm end}\left(\frac{a_{\rm re}}{a_{\rm end}}\right)^{-3(1+\wre)}=\rho^{\rm com}_{\chi}\left(\frac{a_{\rm re}}{a_{\rm end}}\right)^{-4} \implies
    N_{\rm re}=\frac{1}{(1-3\wre)}\text{ln}\left(\frac{\rho^{\rm com}_{\chi}}{\rho_{\rm end}}\right) ,
\end{align}
and the reheating temperature $T_{\rm re}$ is calculated as
\begin{align}\label{Tre1}
  \rho^{\rm com}_{\chi}e^{-4N_{\rm re}}=\frac{\pi^2g_{\rm re}}{30}T_{\rm re}^4 \implies
  T_{\rm re}=\left(\frac{30}{g_{\text{re}}\pi^2}\right)^{\frac{1}{4}}\big(\rho_{\rm end}\big)^{\frac{1}{1-3w_{\phi}}}\big(\rho^{\rm com}_{\chi}\big)^{-\frac{3(1+w_{\phi})}{4(1-3w_{\phi})}} .
\end{align}
Where the background energy density at inflation end is $\rho_{\rm end}=3M_{\rm pl}^2\He^2$,\, $g_{\text{re}}=106.75$ is the total number of relativistic degrees of freedom at the time of reheating in the standard model(SM) for $\Tre \gtrsim 1~\rm GeV$. 
Important to note that as long as the energy spectrum remains blue-tilted, the expressions of both $N_{\rm re}$ and $T_{\rm re}$ in Eq.(\ref{Nre1}) and (\ref{Tre1}) hold true. For any value of coupling strength lying in the range $0\leq\xi<3/16$, we can utilize above two equations for a given comoving energy density $\rho_{\chi}^{\rm com}$ to identify the respective reheating parameters. For example, for $\wre=3/5$, we get the reheating temperatures $\Tre=(6.16\times 10^{-4},~7.95\times 10^{-6})$ GeV for the coupling $\xi=(0, 0.1)$ respectively. Similarly, for $\wre=9/11$, we get $\Tre=(3.83\times 10^{3},~2.43\times 10^{2})$ GeV for the coupling $\xi=(0, 0.1)$ respectively. In this particular range of coupling strength, increasing $\xi$ lowers the temperature as can be recovered from Fig.(\ref{comparisonfig}). As one gradually approaches the conformal limit $\xi=1/6$, large-scale production diminishes substantially, which causes this particular behavior.

\subsection{Reheating parameters $N_{\rm re}, T_{\rm re}$ for $\xi=3/16$}
This is special case for which the number spectrum behaves as $|\beta_k|^2\propto
      (k/k_{\rm end})^{-2\nu_2} $ (see Eq.\ref{no.den2}).
Likewise the previous case, total comoving energy density for $\xi=3/16$ is evaluated to be
\begin{equation}\label{comovingenergy2}
   \rho^{\rm com}_{\chi}\approx \frac{\mathcal{A}_2 \He^4}{4\pi^2\left(2-\nu_2\right)},~~~\mbox{where},~~~~~\mathcal{A}_2=\left(\frac{\Gamma(\nu_2)}{2}\left(\frac{2}{3\mu-1}\right)^{\nu_2}\left|\frac{3\mu-2\nu_2}{4\sqrt{(3\mu-1)}}+\frac{i\sqrt{3\mu-1}}{\pi}\right|\right)^2 .
\end{equation}
As discussed in the previous case, the energy spectrum being blue-tilted for $\xi=3/16$, here we have the similar expressions of reheating parameters $N_{\rm re}, T_{\rm re}$ with $\rho^{\rm com}_{\chi}$ in Eq.(\ref{comovingenergy2}).
\begin{align}\label{Nre2Tre2}
    &N_{\rm re}=\frac{1}{(1-3\wre)}\text{ln}\left(\frac{\rho^{\rm com}_{\chi}}{\rho_{\rm end}}\right)\nonumber\\
    &T_{\rm re}=\left(\frac{30}{g_{\text{re}}\pi^2}\right)^{\frac{1}{4}}\big(\rho_{\rm end}\big)^{\frac{1}{1-3w_{\phi}}}\big(\rho^{\rm com}_{\chi}\big)^{-\frac{3(1+w_{\phi})}{4(1-3w_{\phi})}}
\end{align}
For $\xi=3/16$, we get $\Tre=(3.8\times 10^{-5},~8.96\times 10^2)$GeV for $\wre=(3/5,~ 9/11)$ respectively.

\subsection{Reheating parameters $N_{\rm re}, T_{\rm re}$ for $\xi>3/16$}
For this case, there exists a critical coupling strength $\xi_{\rm cri}(\wre)=\frac{(9 \wre +7) (15 \wre+1)}{48 (3 \wre-1)}$, which  demarcates two regions with a junction at $\xi=\xi_{\rm 
 cri}$. For $\wre>1/3$, the energy-density spectra, $k^4|\beta_k|^2$ becomes perfectly scale-invariant at this critical value. For any $\xi < \xi_{\rm cri}$, the energy spectra remain blue-tilted and turn into red-tilted (or IR divergent) in the regime $\xi > \xi_{\rm cri}$ (Detailed discussion can be found in our earlier paper \cite{Chakraborty:2024rgl}). In this range, we shall compute the reheating parameters in three different regimes.
\subsubsection{For $3/16<\xi<\xi_{\rm  cri}$ :}
For $3/16<\xi<\xi_{\rm cri}$, the energy spectrum behaves as $k^4|\beta_k|^2\propto (k/k_{\rm end})^{4-2\nu_2}$,
and the comoving energy density is calculated to be 
\begin{equation}\label{comovingenergy3}
     \rho^{\rm com}_{\chi}\approx \frac{\mathcal{A}_3 \He^4}{4\pi^2\left(2-\nu_2\right)} 
     \end{equation}
     where
     \begin{equation}
       \mathcal{A}_3\approx\left(\frac{\Gamma(\nu_2)\text{exp}(-\pi \tilde{\nu}_1/2)}{4(3\mu-1)^{-1/2}}\left(\frac{2}{3\mu-1}\right)^{\nu_2}\left|\frac{\left(\pi+i\text{cosh}(\pi\tilde{\nu_1})\Gamma(1-i\tilde{\nu_1})\Gamma(i\tilde{\nu_1})\right)}{\pi\Gamma(i\tilde{\nu_1})}\right|\right)^2  .
     \end{equation}
The corresponding reheating parameters become,
\begin{align}\label{Nre3Tre3}
   &N_{\rm re}=\frac{1}{(1-3\wre)}\text{ln}\left(\frac{\rho^{\rm com}_{\chi}}{\rho_{\rm end}}\right),\nonumber\\
    &T_{\rm re}=\left(\frac{30}{g_{\text{re}}\pi^2}\right)^{\frac{1}{4}}\big(\rho_{\rm end}\big)^{\frac{1}{1-3w_{\phi}}}\big(\rho^{\rm com}_{\chi}\big)^{-\frac{3(1+w_{\phi})}{4(1-3w_{\phi})}} .
\end{align}
For example, for $\wre=3/5$, we find $\Tre=(9.94\times 10^{-3},~5.48\times 10^{-2})$ GeV, and for $\wre=9/11$, we find $\Tre=(9.13\times 10^{4},~5.66\times 10^{5})$ GeV for $\xi=(2, 2.5)$ respectively. Likewise, growing instability with growing $\xi$ causes the increase in temperature in this range of coupling strength also.

\subsubsection{For $\xi=\xi_{\rm cri}$ :}
At this junction point, total energy density is computed as
\begin{align}\label{comovingenergy4}
    \rho^{\rm com}_{\chi}= \frac{\mathcal{A}_3 \He^4}{2\pi^2}\text{ln}\left(\frac{\ke}{\kre}\right) \simeq \frac{\mathcal{A}_3 \He^4(1+3\wre)}{4\pi^2}N_{\rm re} .
\end{align}
Subject to this total energy density, we get the following equation of $N_{\rm re}$, 
\begin{equation}\label{Nre4}
  \left(\text{exp}\left(N_{\rm re}(1-3\wre)\right)-\frac{\mathcal{A}_3 \He^4(1+3\wre)}{4\pi^2\rho_{\rm end}} N_{\rm re}\right) =0 .
\end{equation}
We numerically solve the above equation to find the root $N_{\rm re}$ for a given $\wre$. Using Eq.(\ref{comovingenergy4}) and the solution of the Eq.(\ref{Nre4}), reheating temperature is calculated to be
\begin{equation}\label{Tre4}
T_{\rm re}=\left(\frac{30}{g_{\text{re}}\pi^2}\right)^{\frac{1}{4}}\left( \rho^{\rm com}_{\chi}\right)^{\frac{1}{4}}\text{exp}(-N_{\rm re}) .
\end{equation}
We calculate this critical coupling $\xi_{\rm cri}\simeq(3.23,~2.73)$ for $\wre=(3/5,~9/11)$ respectively. At this critical point, we find the temperatures $\Tre=(13.83,~3.14\times 10^6)$ GeV for the two given EoS. 

\subsubsection{For $\xi>\xi_{\rm cri}$ :} 
After crossing the critical coupling or the junction $\xi_{\rm cri}$, the energy spectrum turns out to be red-tilted. The larger the $\xi$, the heavier the red-tilting of the energy spectrum. 
 Therefore, large scales are now dominantly contributing to the total energy density of the system. Total energy density is now computed as
\begin{align}\label{comovingenergy5}
   &\rho^{\rm com}_{\chi}= \frac{\mathcal{A}_3 \He^4}{4\pi^2(\nu_2-2)}\left(\frac{\ke}{\kre}\right)^{2\nu_2-4}  \simeq \frac{\mathcal{A}_3 \He^4}{4\pi^2(\nu_2-2)}\text{exp}\Big((1+3\wre)(\nu_2-2)N_{\rm re}\Big)
    \end{align}
For this energy density, the reheating e-folding number is calculated to be
\begin{align}\label{Nre5}
   N_{\rm re}=\frac{\text{ln}\left((4\pi^2(\nu_2-2)\rho_{\rm end})/\mathcal{A}_3 \He^4\right)}{\nu_2(1+3\wre)-3(1+\wre)} 
\end{align}
The reheating temperature associated with the  energy density (\ref{comovingenergy5}) and e-folding number (\ref{Nre5}) now becomes
\begin{align}\label{Tre5}
   T_{\rm re}=\left(\frac{30}{g_{\text{re}}\pi^2}\right)^{\frac{1}{4}}\left(\frac{\mathcal{A}_3 \He^4}{4\pi^2(\nu_2-2)}\right)^{\frac{1}{4}}\left(\frac{4\pi^2\rho_{\rm end}(\nu_2-2)}{\mathcal{A}_3 \He^4}\right)^{\frac{(1+3\wre)(\nu_2-2)-4}{4\big(\nu_2(1+3\wre)-3(1+\wre)\big)}}
\end{align}

In this red-tilted regime of the energy spectrum(IR divergent energy spectrum), we get a high reheating temperature. With the increase of the coupling $\xi$, the reheating temperature rises rapidly and reaches the maximum limit($\Tre\sim 10^{15}$GeV) for a coupling strength $\xi\lesssim 50$ as obvious in Fig.(\ref{comparisonfig}). For example, for $\wre=3/5$, we find $\Tre=(3.24\times 10^3,~4.5\times 10^{6})$ GeV, and for $\wre=9/11$, we get $\Tre=(3.36\times 10^{9},~4.5\times 10^{10})$ for $\xi=(3.5, 4)$ respectively.\\
Thus, using all these expressions of $T_{\rm re}$, we study the variation of reheating temperature with respect to the non-minimal coupling strength in this \textit{infrared gravitational reheating scenario}.\\
To this end, we would like to point out one subtle point regarding the consideration of the time-averaged EoS (\ref{eq:wphi}), ignoring the post-inflationary inflaton oscillation while computing the fluctuation spectra. In this work, we propose a special kind of reheating scenario, solely facilitated by the large-scale fluctuations, $k<<\ke$. These long-wavelength modes are subject to strong post-inflationary super-horizon instability beyond a certain coupling strength. Upon horizon reentry at some later time during reheating, they contribute to the total radiation energy density. These scales, having very long wavelengths, cannot sense the small-scale background oscillations after horizon reentry, as the oscillatory feature gradually becomes insignificant at the late stage of reheating. Furthermore, with the increase of the potential exponent $n$, the oscillation time-period of the background also increases, $T\propto \omega^{-1}\propto m_{\phi}^{-1}$(see Eq.(\ref{fre})). So, the background oscillation feature becomes less prominent for higher EoS. Our study reveals that even for a small nonminimal coupling, much smaller than that required for the small-scale resonance instability($\xi>50$) as shown in \cite{Garcia:2023qab}, the substantial growth of these large scales due to super-horizon \textit{tachyonic instability} can successfully reheat the universe. In the subsequent discussion, we shall constrain this reheating dynamics by the CMB scale GW and isocurvature bounds within a very narrow range of $\xi$ values. The large CMB scale will also remain unaffected by the background oscillation. Therefore, for these very long-wavelength classical modes, the consideration of the time-averaged EoS is justified.

\section{Comparing perturbative and infrared reheating}\label{sec4}
In the previous section, we have calculated the reheating parameters $\Tre$ and $\Nre$ taking into account the contribution from the infrared modes which are produced during inflation. However, the usual approach to reheating is
to simultaneously solve the Boltzmann equations for decaying inflaton and radiation. Such an approach naturally deals with the excitation of sub-Hubble modes \cite{Chakraborty:2025zgx} and their subsequent thermalization at the end of reheating.   
In this section, we shall discuss the growth of sub-horizon modes using the standard perturbative technique, i.e., the Boltzmann approach, and compare its contribution with that of the super-horizon components discussed in the previous section. In order to do this, we need to track the evolution of the inflaton $(\rphi)$ and the radiation ($\rhor$) energy densities during reheating, we solve the following set of coupled Boltzmann equations 
\begin{eqnarray}
 \dot{\rho_{\rm \phi}}+3H(1+w_\phi)\rho_\phi= -Q ,  \label{Boltzman1}~~;~~
\dot{\rhor}+4H\rhor=Q,~~;~~
\label{Boltzman2}
H^2=\frac{\rho_{\rm \phi}+\rhor}{3\,M_{\rm pl}^2}\label{Boltzman3}\,,
\end{eqnarray}
where $Q$ is the production rate, the amount of energy transferred per unit time, and the unit volume. Since the inflaton primarily governs the total energy density during the reheating period, the expansion rate associated with the term $3H\,(1+\wphi)\,\rphi$ significantly surpasses the reaction rate $Q$. As a result, the inflaton part of Eq.(\ref{Boltzman1}) can be solved analytically by disregarding the right-hand side, leading to the solution,
\begin{equation}{\label{rhophi}}
    \rphi(a)\simeq\rho_{\rm end}\left(\frac{a}{\aend}\right)^{-3(1+\wphi)},
\end{equation}
with corresponding Hubble rate
\begin{equation}\label{eq:Hubble}
    H(a)\simeq H_{\rm end}\left(\frac{a}{\aend}\right)^{-\frac{3}{2} (1+\wphi)} .
\end{equation}
In the present context, we consider a non-minimally coupled scalar field ($\chi$) as radiation discussed earlier. The total production rate must be the sum of the minimal and the non-minimal gravitational interactions. To this end, \textit{it is important to note that for a non-minimally coupled theory, frame ambiguity exists in the computation of the perturbative production rate, and this is what we analyze now in the following discussion}.

By utilizing the conformal transformation, the production rate of massless non-minimally coupled scalar field $\chi$ has already been computed \cite{Co:2022bgh, Barman:2022qgt} in the Einstein frame and studied their implication in the reheating dynamics. However, note that in the super-horizon analysis as discussed in the previous section, one usually solves for the $\chi$ field in the Jordan frame. Hence, in order to compare with the results we obtained for the infrared modes, we shall compute the scalar field production rate in the Jordan frame. For completeness, we also analyze the difference in predictions for both frames. Any dissimilarity in the outcomes will indicate the fundamental difference in these frame transformations.
The total production rate expressions in two frames (see the detailed derivation in the Appendix \ref{appenproductionrate}) are given by,
\begin{equation}{\label{rk}}
    Q\simeq\frac{\rphi^2\,m_\phi}{8\,\pi\,M_{\rm pl}^4}\,\mathcal S^{\rm\xi}_{\rm n}\,,
\end{equation}
where given the inflaton potential Eq.(\ref{alphapotential}), $\mathcal S^{\rm\xi}_{\rm n}$ are time independent constant, and the expressions are,
\begin{align}\label{eq:EJrate}
    \mathcal{S}^{\rm\xi}_{\rm n}=
    \begin{cases}
    \bigg((1-6\xi)^2\sum^{\infty}_{\nu=1}\,\gamma\,\nu\,\lvert\mathcal P^{ 2n}_\nu\rvert^2\bigg)\quad & \text{Jordon frame}\\\\
\bigg(\underbrace{\sum^{\infty}_{\nu=1}\gamma\,\nu\,\lvert\mathcal P^{ 2n}_\nu\rvert^2}_{\text{for minimal case}}\bigg)+\bigg(\underbrace{\,\xi^2\sum^{\infty}_{\nu=1}\,\gamma\,\nu\,\left\lvert2\,\mathcal{P}^{ 2n}_\nu+\frac{2n\,(2n-1)\,\gamma^2\,\nu^2}{2}\lvert\mathcal P_\nu\rvert^2\right\rvert^2}_{\text{additional term for non-minimal case}}\bigg)\quad & \text{Einstein frame}
\end{cases}
\end{align}
While evaluating the interaction rate, the inflaton is treated
as a time-dependent external and classical background field, which we parametrize as
\begin{equation}\label{eq:phi}
    \phi(t)=\phi_0(t).\mathcal{P}(t)=\phi_0(t)\sum_\nu \mathcal{P}_\nu e^{-i \nu \omega t}\,,
\end{equation}
where  $\phi_0(t)$ represents the decaying amplitude of the oscillation and $\mathcal{P}(t)$ encodes the oscillation of the inflaton with the fundamental frequency calculated to be \cite{Garcia:2020wiy}
\begin{equation} \label{fre}
\omega = m_\phi(t)\,\gamma\,,~~~\mbox{where}~~\gamma=\sqrt{\frac{\pi\,n}{2\,n-1}}\frac{\Gamma\left(\frac{1}{2}+\frac{1}{2\,n}\right)}{\Gamma\left(\frac{1}{2\,n}\right)},\,m_{\phi}\rightarrow\mbox{mass of inflaton} 
\end{equation}
It is interesting to note the important difference in decay rates of inflaton, or in other words, the production rate of the conformally coupled $\chi$ field in both frames. The noticeable difference can be observed at $\xi = 1/6$. The Jordan frame production rate vanishes at this special point, as expected due to the conformal property of the massless scalar field. On the other hand, in Einstein's frame, such a property ceases to exist, which is also reflected in the above formula, yielding non-vanishing production.

Utilizing Eq.(\ref{rhophi}) along with the interaction rate Eqs.(\ref{rk}) and (\ref{eq:EJrate}) in Eq.(\ref{Boltzman2}), one can obtain the radiation energy density as follows
\begin{equation}\label{rhorg}
    \rhor(a)\simeq\mathcal{S}^\xi_{\rm n}\frac{9\,H^3_{\rm end}\,m^{\rm end}_{\rm\phi}}{4\,\pi\,(1+15\,w_{\rm\phi})}\left(\frac{\aend}{a}\right)^{4}\left[1-\left(\frac{a}{\aend}\right)^{-\frac{
    1+15\wphi}{2}}\right],
\end{equation}
where $m^{\rm end}_{\rm\phi}$ is the inflaton mass at the end of inflation.
In the presence of the  non-minimal coupling, therefore, the reheating temperature $\Tre$ and reheating e-folding number $N_{\rm re}$ can be expressed for any arbitrary value of the $\xi$ as,
\begin{equation}
 \tre\simeq0.5\MP\left(\frac{3\,\mathcal{S}^\xi_{\rm n}}{4\pi(1+15w_{\rm\phi}) }\frac{m^{\rm end}_{\rm\phi}}{\mpl}\right)^{\frac{3(1+w_\phi)}{4(3w_{\rm\phi}-1)}}\left(\frac{H_{\rm end}}{\mpl}\right)^{\frac{9w_\phi+1}{4(3w_\phi-1)}}
     \label{gretre}
 \end{equation}
 \begin{equation}
     N_{\rm re}=\frac{1}{(3\,\wphi-1)}\,\text{ln}\left[\left(\frac{4\,\pi\,(1+15\,\wphi)}{3\,\mathcal{S}^\xi_{\rm n}}\right)\left(\frac{\MP}{\hend}\right)\left(\frac{\MP}{\mphi}\right)\right]
 \end{equation}
For the varying non-minimal coupling associated function $\mathcal{S}_{\rm n}^{\xi}$ in two frames, we obtain different predictions of reheating parameters, specifically around the conformality, $\xi\sim1/6$. As pointed out earlier, since $\mathcal{S}^\xi_{\rm n} \propto (1-6\xi)$ in the Jordan frame, the radiation temperature clearly vanishes, as there is no radiation production as opposed to the Einstein frame temperature. This can also be observed from our full numerical computation depicted in Fig.(\ref{comparisonfigJorEin}) in Appendix \ref{appenJorEin}. In Fig.(\ref{comparisonfigJorEin}), it is seen that in the higher non-minimal coupling regime, $\xi>1$, pertinent to the present analysis, frame ambiguity is gone. Therefore, in the subsequent discussion, we shall stick to the Jordan frame while comparing the perturbative and non-perturbative reheating parameter predictions. 


\subsection{Comparing perturbative and non-perturbative approaches in infrared gravitational production}
\begin{figure}[t]
     \begin{center}
\includegraphics[scale=0.45]{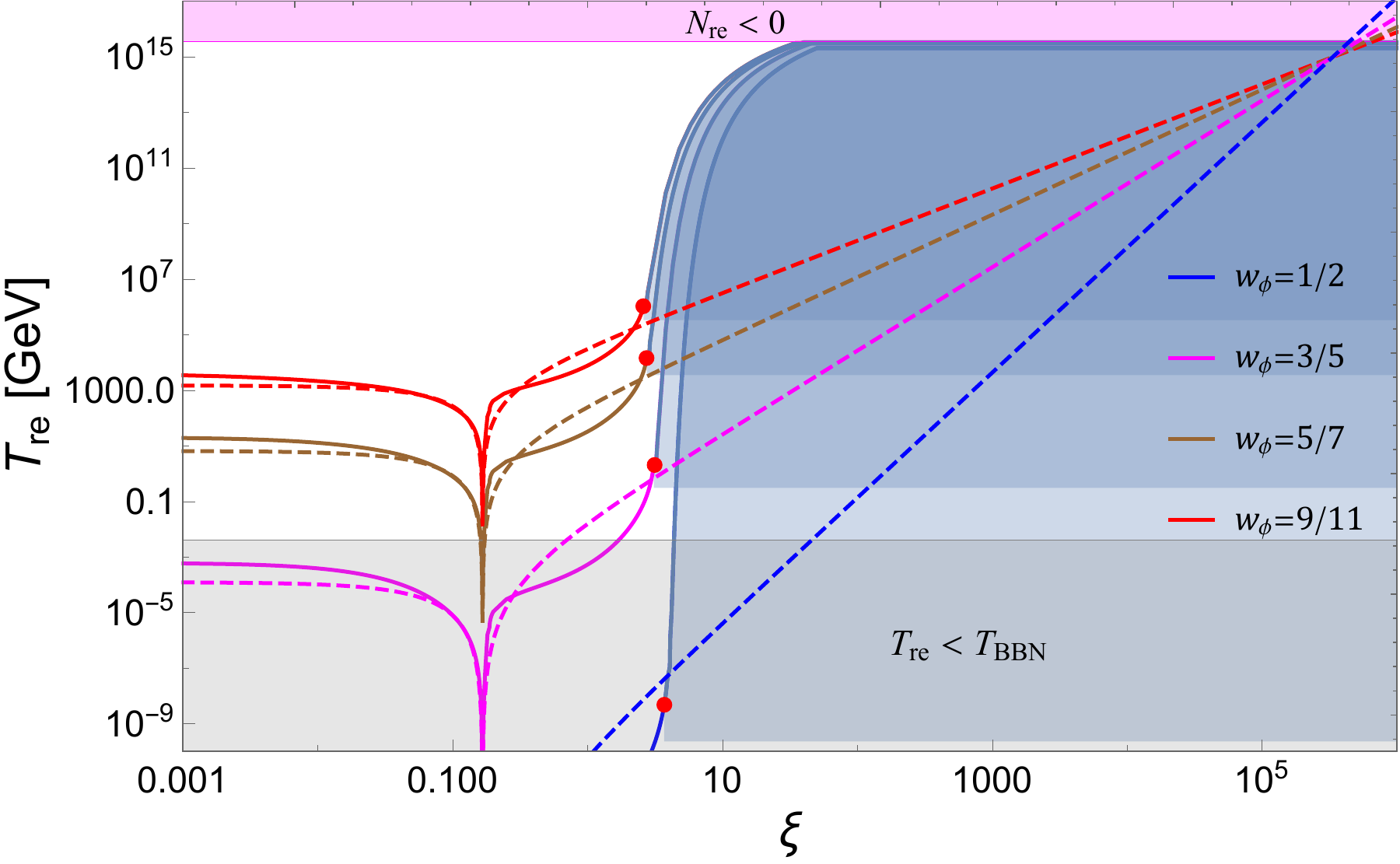}
\caption{\textit{ Figure represents a comparison of $T_{\rm re}$ vs $\xi$ variation for different EoS between perturbative(Jordan frame) and non-perturbative analysis. Solid lines correspond to the non-perturbative or Bogoliubov predictions, and dashed lines correspond to the perturbative Boltzmann predictions in the Jordan frame.
Red dots on each color line indicate the maximum value of the coupling strength, $\xi_{\rm max}$, corresponding to each EoS consistent with large-scale observational bounds ( gravitational wave and isocurvature), and the overlapping shaded region is ruled out by the latest CMB scale tensor-to-scalar ratio and isocurvature bounds for all four EoS. For a given EoS, any $\xi$, exceeding the red dot is disallowed by observation, and its associated $\Tre$ lies inside the shaded region. There are four demarcating lines inside the shaded region. From the top, the first one corresponds to $\wre=9/11$, the second one to $\wre=5/7$, the third one to $\wre=3/5$, and the fourth one to $\wre=1/2$. 
}} 
\label{comparisonfig}
\end{center}
\end{figure}

Non-minimal coupling produces large infrared fluctuations via tachyonic instability. After the end of inflation, on the other hand, the perturbative effect (effect of small scales) on reheating cannot be ignored. To compare the contribution of these two production processes, we further realized that it is appropriate to consider the Jordan frame. Hence, in this frame, assuming both the contributions as independent in Fig.(\ref{comparisonfig}), we have shown how non-perturbative infrared predictions of reheating temperature surpass that of the perturbative prediction beyond a certain coupling strength. 
These takeover happen around $\xi\approx(4,~2.9,~2.4,~2.2)$ for $\wre=(1/2,~3/5,~5/7,~9/11)$, respectively(see also Table \ref{tab1} showing reheating temperature predictions for three EoS). 
Therefore, the usual perturbative approach as discussed in \cite{Co:2022bgh, Barman:2022qgt} can be observed to greatly underestimate the value of reheating temperature as depicted in dotted lines as compared to the infrared contribution depicted in solid lines in Fig.(\ref{comparisonfig}). For example, it requires as large as $\xi \simeq 10^5$ to reach the maximum reheating temperature $T_{\rm max}\simeq 10^{15}$ GeV in the perturbative approach as indicated by the point where all the dashed-colored lines meet in Fig.(\ref{comparisonfig}). On the contrary, a value as small as $\xi \lesssim 50$ turned out to be sufficient to obtain maximum reheating temperature from the super-horizon modes, as can be seen from the converging solid color lines. 

To this end, let us briefly state the peculiar feature (dominance of perturbative prediction over the non-perturbative) observed in  Fig.(\ref{comparisonfig}), particularly in the coupling range $1/6<\xi<\xi_{\rm cri}$. As stated earlier, 
infrared instability is expected to be suppressed around the conformal value $\xi\gtrsim 1/6$, and becomes stronger only for larger values of $\xi \gtrsim\xi_{\rm cri}$, beyond which the energy spectrum becomes red-tilted. The moment the spectrum becomes red tilted, one naturally expects the total energy budget to be dominated by the super-horizon mode. This is when all the solid curves corresponding to the contribution from the infrared modes start to show an upward trend with increasing $\xi >\xi_{\rm cri}$ becoming stiff, surpassing the contribution from that of the perturbative sub-horizon contribution.


Our analysis so far, therefore, indicates that by increasing the $\xi$ value, we can increase the radiation temperature as high as $10^{15}$ GeV, and the infrared modes solely contribute to that high reheating temperature. Note that the origin of such a high temperature is due to the infrared divergent energy spectrum, particularly for $\wre >1/3$.  
However, in the subsequent sections, we shall see that from the observational perspective, this IR divergent energy spectrum, particularly at the large (CMB) scales, often becomes problematic and is tightly constrained by observational bounds, like the tensor-to-scalar ratio and the isocurvature constraint.
We now derive the constraints on $\xi$ in light of the latest \textit{Planck}, ACT, DESI, and
 BICEP/Keck combined (P+ACT+LB+BK18) bound on tensor-to-scalar ratio and isocurvature at the CMB scale provided in \cite{ACT:2025tim, ACT:2025fju}. The constraints on $\xi$ will, anyway, further constrain the present infrared reheating dynamics. This is the subject of our subsequent discussion.
\begin{table}[t]
\caption{\textit{Variation of reheating temperature with  EoS and non-minimal coupling strength for $\alpha=1$. Here we have given the perturbative and non-perturbative predictions for comparison. The Jordan(Einstein) frame predictions are given without and with braces in the perturbative analysis. 
}}
 \vspace{2ex}
 \centering
\begin{tabular}{|c|c|c|c|c|}
\hline
\multicolumn{1}{|c|}{EoS ($\wre$)} & \multicolumn{2}{|c|}{Non-perturbative} & \multicolumn{2}{|c|}{Perturbative}  \\
\hline
\multirow{5}{*}{3/5}& $\xi$& $T_{\text{re}}~(\text{GeV})$ &  $\xi$ & $T_{\text{re}}~(\text{GeV})$\\
\cline{2-5}
   & 0& $6.16\times10^{-4}$& 0& $6\times10^{-4}$\,($6\times10^{-4}$) \\
   \cline{2-5}
   & 3/16& $3.8\times 10^{-5}$& 3/16& $9.38\times10^{-6}$\,($2.0\times10^{-3}$) \\
   \cline{2-5}
   & 1& $5.3\times 10^{-4}$& 1& $7.5\times10^{-2}$\,($1.3\times10^{-1}$) \\
   \cline{2-5}
   \cline{2-5}
   & $ \xi_{\rm max}=2.95$& $3.3$& $\xi_{\rm max}=2.95$& $2.8$\,(2.7) \\
   \cline{2-5}
   & $\xi_{\rm cri}(3/5)$& 13.84& $\xi_{\rm cri}(3/5)$& 3.73\,($3.0$) \\
\hline
  \multirow{5}{*}{5/7}& 0& 20.35 & 0& 6.5\, ($6.5$) \\
 \cline{2-5}
   & 3/16& 3.14& 3/16& $2.87\times 10^{-1}$\, (17) \\
   \cline{2-5}
   & 1& 32.4& 1& $2.43\times 10^2$\,($4.0\times10^2$) \\
   \cline{2-5}
   & $ \xi_{\rm max}=2.63$& $5.63\times 10^3$& $ \xi_{\rm max}=2.63$& $2.78\times 10^3$\,($2.6\times10^3$) \\
   \cline{2-5}
   & $\xi_{\rm cri}(5/7)$& $4.93\times 10^4$& $\xi_{\rm cri}(5/7)$& $3.43\times 10^3$\,($3.0\times10^3$) \\
   \cline{2-5}
  \hline
  \multirow{5}{*}{9/11}& 0& $3.83\times 10^3$ & 0&\,$2.0\times10^3$\, ($2.0\times10^3$)  \\
 \cline{2-5}
   & 3/16& $8.96\times 10^2$& 3/16& $1.48\times 10^2$\, ($3.0\times10^3$) \\
   \cline{2-5}
   & 1& $7.68\times 10^3$& 1& $4.09\times 10^4$\,($4.0\times10^4$) \\
   \cline{2-5}
   & $ \xi_{\rm max}=2.51$& $6\times 10^5$& $ \xi_{\rm max}=2.51$& $2.84\times 10^5$\,($10^5$) \\
   \cline{2-5}
   & $\xi_{\rm cri}(9/11)$& $3.14\times 10^6$& $\xi_{\rm cri}(9/11)$& $3.37\times 10^5$\,($2.0\times10^5$) \\
 \hline
\end{tabular}
\label{tab1}
\end{table}

\section{Constraints from Gravitational Wave and Isocurvature Perturbation \label{secgw}
}
Gravitational waves (GWs) probe very deep in the early universe due to their extremely weak coupling with matter fields. Therefore, any early universe model that can generate GW can, in principle, be constrained through direct or indirect observation.   

\subsection{Constraining the infrared reheating dynamics through tensor-to-scalar ratio $r_{0.05}$:}

 For stiff EoS $\wre>1/3$, a stronger IR instability (tachyonic instability) of the scalar fluctuations generates larger tensor fluctuations even at the Cosmic Microwave Background(CMB) scale beyond a certain threshold of $\xi$. In this discussion, we shall closely follow the reference \cite{Chakraborty:2024rgl} to write down the expression of the secondary tensor power spectrum being sourced by these scalar fluctuations. For very long-wavelength modes $k<<\kre$, the secondary tensor power spectrum is written as\cite{Chakraborty:2024rgl}
\begin{align}\label{tensorpower}
\lim_{k<<\kre}\Pt^{\rm sec}(k,\ere)\simeq &  \frac{2\mathcal{A}_3^2\He^4}{\pi^4\mpl^4}\l\{ \frac{1}{2l(\delta-2)}+ \frac{1}{4l(1-l)-2l\delta}\r\}^2 \frac{8(1+2\nu_2)}{15(3-\nu_2)(4\nu_2-5)}\nno\\
&\times \l( \frac{\ke}{\kre}\r)^{4-2\delta}\l( \frac{k}{\ke}\r)^{4(2-\nu_2)} 
\end{align}
 Using the above amplitude of tensor fluctuations (\ref{tensorpower}), and assuming that maximum contributions originate from secondary sources (i.e., the scalar fields), we obtain the tensor-to-scalar ratio at the CMB pivot scale as
\begin{align}\label{tensortoscalar}
r_{0.05}&\simeq \frac{2\mathcal{A}_3^2\HI^4}{\pi^4\mpl^4 A_s}\l\{ \frac{1}{2l(\delta-2)}+ \frac{1}{4l(1-l)-2l\delta}\r\}^2 \frac{8(1+2\nu_2)}{15(3-\nu_2)(4\nu_2-5)}\nn\\
    &\times 
    e^{\l(\Nre(6\wre-2)\r)}\l( \frac{k_{\ast}}{\ke}\r)^{4(2-\nu_2)}\leq 0.038
\end{align}

where the above parameters are expressed in terms of $\wre$ and $\xi$ as 
$l=3(\wre-1)/2(1+3\wre),~ \delta=4/(1+3\wre)$. The ratio $\l(\ke/\kre\r)$ is written as $\l(\ke/\kre\r)=\text{exp}\l(\Nre(1+3\wre)/2\r)$. Here \enquote{$k_{\ast}$} is the 
pivot scale or CMB scale, $(k_{\ast}/a_0)=0.05\text{Mpc}^{-1}$ and $a_0$ is the present scale factor. According to the combined(P+ACT+LB+BK18) data set, there is no significant improvement in the tensor-to-scalar ratio bound, $r_{0.05}<0.038$ at the 95$\%$ C.L \cite{ACT:2025tim}. This slight modification of the maximum $r_{0.05}$ bound will not relax the upper limit on $\xi$ much. 
In the background of the present reheating scenario, from the above inequality (\ref{tensortoscalar}), for a given reheating EoS $\wre>1/3$,  we obtain a maximum value of coupling strength $\xi_{\rm max}$ to prevent the overproduction of tensor fluctuations at the CMB scale. Any coupling strength $\xi>\xi_{\rm max}$, represented by the red dots in the Fig.(\ref{comparisonfig}), is discarded by the current tensor-to-scalar ratio bound at the CMB scale, which is depicted by the blue shaded region. This upper boundary of $\xi$ in turn sets the maximum allowed reheating temperature for a given EoS in the present reheating scenario. 
\subsection{Constraints from isocurvature power spectrum $\Ps(k_{\ast})$ : }
As we have observed, massless scalar long-wavelength modes experience substantial post-inflationary growth, being driven by the super-horizon instability induced by the non-minimal gravity coupling as previously illustrated. This large-scale instability inevitably generates significant isocurvature fluctuations at the CMB scale. The current constraint on the isocurvature power spectrum by \textit{Planck} 2018
is defined to be $\beta_{\rm iso}\equiv \Ps(k_{\ast})/\l(\Pr(k_{\ast})+\Ps(k_{\ast})\r)\lesssim 0.038$ at the $95\%$ C.L for the 
 pivot scale or CMB scale $k_{\ast}$\cite{Planck:2018jri}. The pivot scale amplitude of curvature power spectrum $\Pr(k_{\ast})=2.1\times 10^{-9}$ gives the upper bound of the amplitude of isocurvature power spectrum at CMB scale $\Ps(k_{\ast})\lesssim 8.3\times 10^{-11}$\cite{Planck:2018jri, Planck:2018vyg}. 
 This large-scale upper bound of the isocurvature power spectrum constrains the reheating dynamics further.\\  The second-order isocurvature power spectrum is evaluated by using the following expression as \cite{Garcia:2023awt, Garcia:2023qab, Chung:2004nh, Chung:2011xd, Ling:2021zlj, Kolb:2023ydq, Liddle:1999pr} 
\begin{equation}\label{isocurvpow}
  \Ps(k)= \frac{1}{\rho_{\chi}^2}\frac{k^3}{2\pi^2}\int d^3\vec{x}\langle\delta\rho_{\chi}(\vec{x})\delta\rho_{\chi}(0)\rangle e^{-i \vec{k}.\vec{x}}=\frac{k^3}{(2\pi)^5 \rho_{\chi}^2a^8}\int d^3\vec{p}~ P_{X}\l(p,|\vec{p}-\vec{k}|\r) 
\end{equation}
where $\rho_{\chi}$ and $\delta\rho_{\chi}$ are energy-density and its fluctuation of the field having finite mass $m_{\chi}$, and the functional integrand for non-zero mass is given by 
\begin{equation}\label{PX1}
    P_{X}(p,q)=|X_p^{\prime}|^2|X_q^{\prime}|^2+a^4m_{\chi}^4|X_p|^2|X_q|^2+a^2m_{\chi}^2 \l[(X_p X_p^{\prime\ast})(X_q X_q^{\prime \ast})+h.c\r] .
\end{equation}
As here we are dealing with purely massless fields, only the non-vanishing contribution will come from the first term of the expression of $P_{X}(p,q)$ above, and similarly, $\rho_{\chi}$ is the energy-density of the massless scalar. To evaluate this integral, we shall exploit the late-time post-inflationary solutions of the rescaled field mode $X_k$ in different ranges of $\xi$ values as found in \cite{Chakraborty:2024rgl}. From our earlier discussion in Section \ref{sec2}, 
we only confine ourselves to the region where $\wre>1/3$, $\xi>3/16$ \cite{Chakraborty:2024rgl}. 

\begin{figure}[t]
\begin{center}
\includegraphics[scale=0.27]{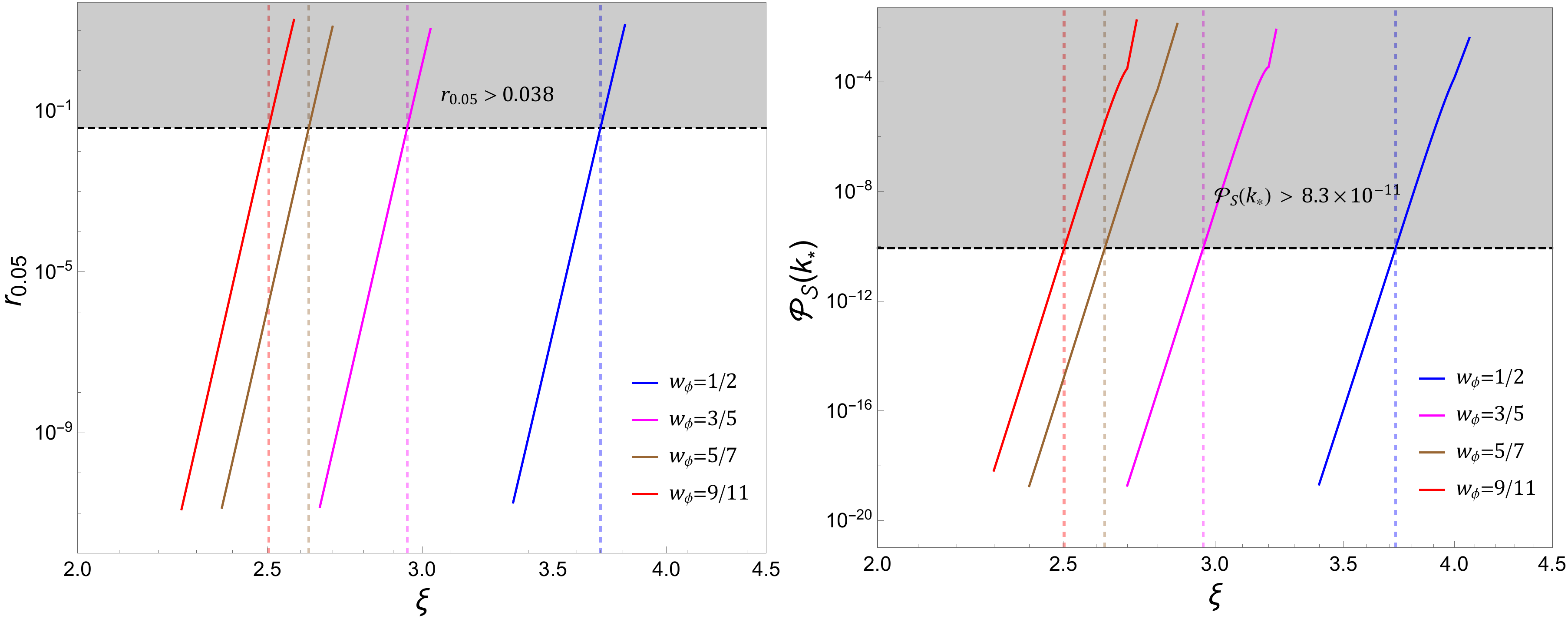}
\caption{\textit{\textbf{Left Panel}: 
Figure represents constraints on $\xi$ for different post-inflationary EoS. The horizontal black dashed line indicates the maximum bound on $r$ at the CMB scale, $r_{0.05}=0.038$, and the vertical colored dashed lines show the maximum allowed value of the coupling $\xi_{\rm max}$ for different EoS subject to this bound. For a given EoS, $r_{0.05}$ for any $\xi>\xi_{\rm max}$(shaded region) is hence disallowed by the current bound.
 \textbf{Right panel}: Figure represents the constraints on $\xi$ from isocurvature power spectrum for different post-inflationary EoS. The horizontal black dashed line indicates the current bound on CMB scale isocurvature amplitude $\Ps(k_{\ast})=8.3\times 10^{-11}$, and the vertical colored dashed lines show the maximum allowed value of $\xi$ for different EoS. For a given EoS, $\Ps(k_{\ast})$ for any $\xi>\xi_{\rm max}$(shaded region) is hence disallowed by the current isocurvature bound.
}} 
\label{rmcmbisoampvsxifig}
\end{center}
\end{figure}

For $\wre>1/3$ and $\xi>3/16$, evaluating the integral (\ref{isocurvpow}), we find the expression of the large-scale isocurvature power spectrum for massless fluctuation as follows:
\begin{align}\label{isofinalamp}
\Ps(k)=
\frac{\mathcal{A}_3^2\ke^8 {\cal I}_2}{(2\pi)^4(3-2\nu_2)\rho_{\chi}^2a^8}  \l(\frac{k}{\ke}\r)^{(8-4\nu_2)} .
\end{align}
The detailed computation of the above Eq.(\ref{isofinalamp}) with the functional form of ${\cal I}_2$ is found in Appendix \ref{appenA}. 
Substituting the expressions (\ref{comovingenergy3}) and (\ref{comovingenergy5}) into  (\ref{isofinalamp}) we compute the expressions of the isocurvature power spectrum in the large scale($k<<\ke$) as follows:
\begin{align}\label{isoamp}
  \Ps(k)=&
\l(\frac{k}{\ke}\r)^{(8-4\nu_2)} \l(\frac{2-\nu_2}{\sqrt{3-2\nu_2}}\r)^2 {\cal I}_2
\times
\begin{cases}
     1 ~~~~~~~~~~~~~~~~~~~~~~~~~~\text{for}~ 3/16<\xi<\xi_{\rm cri}\\
    e^{(1+3\wre)(4-2\nu_2)N_{\rm re}} ~~~~~\text{for}~ \xi>\xi_{\rm cri}
\end{cases} .
\end{align}
Interestingly, we have obtained the same spectral index $(8-4\nu_2)$ as it is found in 
tensor power spectrum for $k<<\kre$ in Eq. (\ref{tensorpower}). This confirms a behavioral similarity between isocurvature and tensor power spectrum in the long-wavelength regime ($k<<\kre$) for a given $\wre$ and coupling strength $\xi$.
The CMB constraint on isocurvature spectrum $ \Ps(k_{\ast}) <  8.3\times 10^{-11}$ immediately gives an another upper bound 
on $\xi_{\rm max}$. Our numerical computation shows that both the upper bounds from tensor to scalar ratio, derived from Eq.(\ref{tensortoscalar}), and isocurvature constraints, derived from Eq.(\ref{isoamp}), are approximately the same clearly depicted in 
Fig.(\ref{rmcmbisoampvsxifig}).
The reason behind this can be explained through the 
relation between isocurvature and tensor power spectrum at the CMB scale. 
Using the inequality $r_{0.05}=\Pt(k_{\ast})/\Pr(k_{\ast})\lesssim 0.038$, we can express $\Pr(k_{\ast})=\Pt(k_{\ast})/\l(r_{0.05}\lesssim 0.038\r)$. Substituting this scalar power spectrum amplitude into $\beta_{\rm iso}$, the relation $\beta_{\rm iso}=\Ps(k_{\ast})/\l(\Pr(k_{\ast})+\Ps(k_{\ast})\r)\sim 0.038$ can be translated into the approximate relation $\Ps(k_{\ast})\approx 1.04 \Pt(k_{\ast})$. This relation suggests that these two bounds predict more or less the same maximum bound $\xi_{\rm max}$ for any particular reheating EoS.

Based on these two large-scale observational bounds, we get a very narrow allowed regime in the reheating parameter space, where the non-perturbative prediction 
surpasses the perturbative prediction, as obvious in Fig.(\ref{comparisonfig}). The red dots in Fig.(\ref{comparisonfig}) indicate the $\xi_{\rm max}$ values for four different EoS respecting the relevant observational bounds as already discussed in this Section.  For example, we get the narrow ranges, $2.9\lesssim\xi\lesssim 2.95$, $2.4\lesssim \xi\lesssim 2.63$, and $2.2\lesssim\xi \lesssim 2.51$ for $\wre=3/5,~5/7,~9/11$ respectively, where the non-perturbative prediction slightly overtakes the perturbative one. 
It is in this coupling range, the infrared fluctuations are dominated, and hence one should expect the production of gravitational waves, which can be relevant for the existing and upcoming gravitational waves experiments. 


\section{GW signature of infrared gravitational reheating}\label{secgwsign}
In light of the current CMB scale tensor-to-scalar ratio and isocurvature bound as discussed above, we get a very narrow range of allowed coupling strength for any EoS $\wre>1/3$. And each $\wre$ has a maximum limit of $\xi$ consistent with the current bounds of $\Ps(k_{\ast})$ and
$r_{0.05}\lesssim 0.038$ as discussed. In this section, we study the spectral nature of the total gravitational wave spectrum, combining the primary and secondary parts in all frequency scales.

\textbf{\underline{Behavior of primary gravitational wave (PGW) spectrum today :}}
The PGW spectrum is the one which is originated from quantum fluctuation during inflation. The present-day PGW spectrum are read as \cite{Haque:2021dha, Maiti:2024nhv, Chakraborty:2024rgl, Maiti:2025cbi}:

\begin{align}
    \ogwp(k)h^2\simeq \Omega_Rh^2\l(\frac{\HI ^2}{12 \pi^2\MP^2}\r)\times\l\{\begin{matrix}
        1 & &\mbox{ for}~~ k<\kre\\
       \frac{(1+3\wre)^{(2-\nw)}}{\pi}\Gamma^2\l(\frac{5+3\wre}{2+6\wre}\r) \left(\frac{k}{\kre}\right)^{\nw} & &\mbox{ for}~~ \kre<k<\ke
    \end{matrix}\r. .\label{eq:gws_pri}
\end{align}
where the spectral index \enquote{$\nw$} is defined as $\nw=2(3\wre-1)/(1+3\wre)$, and \(\Omega_{\rm R} h^2 \simeq 4.3 \times 10^{-5}\) denotes the dimensionless energy density of radiation at the current epoch\cite{Planck:2018vyg}.

\textbf{\underline{Behavior of scalar induced secondary gravitational wave(SGW) spectrum today :}}

Likewise, the primary spectra, we can also define the scalar field-induced secondary spectra for today in the regime $\xi>3/16$ as illustrated in \cite{Chakraborty:2024rgl}.
\begin{align}
\ogws(k)h^2=&\frac{\l(1+k^2/\kre^2\r)\mathcal{P}_{\rm T}^{\rm sec}(k,\ere)}{24}\simeq~ \Omega_Rh^2\frac{\mathcal{A}_3^2\HI^4}{12\pi^4\mpl^4}\frac{8(1+2\nu_2)}{15(3-\nu_2)(4\nu_2-5)}\l( \frac{\ke}{\kre}\r)^{4-2\delta}\nonumber\\
&\times
\begin{cases}
\l( \frac{1}{2l(\delta-2)}+ \frac{1}{4l(1-l)-2l\delta}\r)^2 \l(\frac{k}{\ke}\r)^{8-4\nu_2} & \text{for}~~ k<\kre\\
 \frac{2^{1-2l-2\delta}\pi\Gamma^2(1-l)\Gamma^2(l)}{\Gamma^2(l+ \frac{\delta}{2})\Gamma^2\l( \frac{\delta}{2}\r)} \l(\frac{\ke}{\kre}\r)^{\delta-2} \l(\frac{k}{\ke}\r)^{6+\delta-4\nu_2}& \text{for} ~~ \kre<k<\ke
\end{cases}
\label{eq:gws}
\end{align}

where the indices $l,~ \delta$ are functions of $\wre$ as defined earlier.

\begin{figure}[t]
     \begin{center}
\includegraphics[scale=0.295]{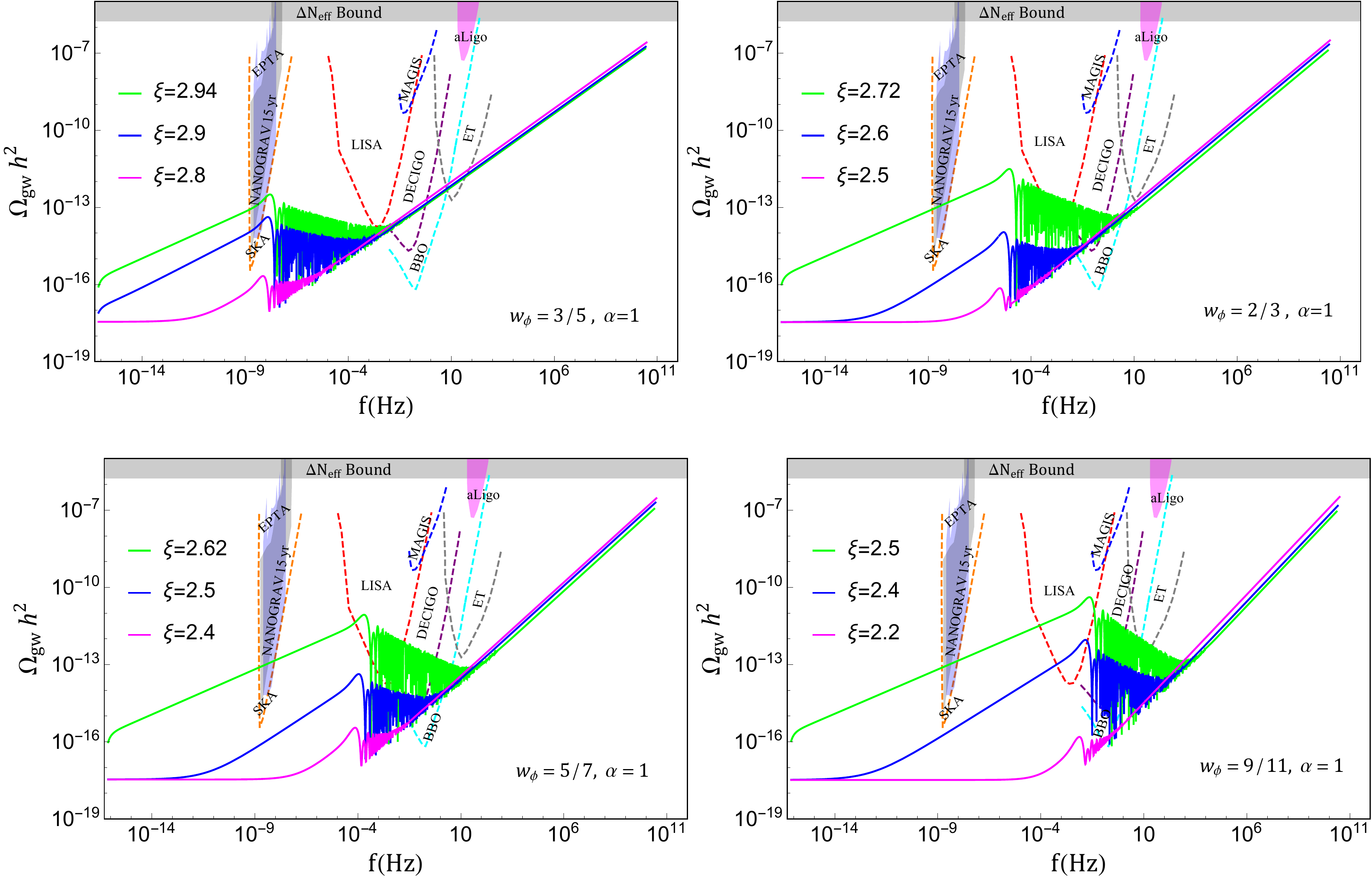}
\caption{\textit{Figure represents the total gravitational wave spectra (PGW+SGW) for today with the variation of non-minimal coupling strength for different reheating EoS in the non-minimal coupling-induced infrared reheating background. For each EoS, a particular coupling $\xi$ exists at which primary strength starts to overcome the secondary one at all frequency regimes, and the total GW spectrum closely follows the primary behavior, as obvious in this figure. For instance, for $\wre=2/3,~ \xi\lesssim 2.5$, PGW strength starts to surpass the secondary strength in the entire frequency scale. In this plot, we have taken the lowest frequency, $f_{\ast}= \l(k_{\ast}/2\pi\r)\sim 7.75\times 10^{-17}$ Hz, and the highest frequency, $f_{\rm end}=\l(\ke/2\pi\r)\sim 10^{11}$ Hz for all the EoS.} } 
\label{gwspectrumfig}
\end{center}
\end{figure}

Combining these two spectra, we define the total gravitational wave spectrum for today, $\Omega_{\rm gw}h^2=(\ogwp(k)h^2+\ogws(k)h^2)$, whose spectral nature is depicted in Fig.(\ref{gwspectrumfig}) in the present reheating background, and they are consistent with the current observational bounds at CMB scales discussed in the sections above. 
In addition, we also take into account $\Delta N_{\rm eff}$ bound on GW energy density $\Omega_{\rm gw}h^2\lesssim 1.7\times 10^{-6}$ \cite{Clarke:2020bil} where generically high-frequency modes contribute. In Fig.(\ref{gwspectrumfig}), in all the plots, one common feature we notice is that at low and intermediate frequency ranges, the secondary strength significantly surpasses the primary one beyond a certain coupling for a given $\wre$. In this coupling regime, the total GW spectrum assumes a broken power law form with three different spectral indices as
\begin{align}
\Omega_{\rm gw}(k)h^2\propto
\begin{cases} 
{k}^{8-4\nu_2} & \text{for}~~ k<\kre\\
{k}^{6+\delta-4\nu_2}& \text{for} ~~ \kre<k<k_{\rm SB} \\
{k}^{\nw} &\mbox{ for}~~ k_{\rm SB} <k<\ke
\end{cases}
\label{break}
\end{align}
Where $k_{\rm SB}$ corresponds to spectral breaking scale at which 
$\ogws(k)h^2=\ogwp(k)h^2$ is satisfied, and 
above this, the primary GW spectrum dominates. 
Equating the Eq. (\ref{eq:gws}) with  (\ref{eq:gws_pri}) and simplifying we get this characteristic frequency as
\begin{align}\label{eq:kSB}
 \left(\frac{k_{\rm SB}}{\ke}\right)=\left(\mathcal{C}~~ \text{exp}\left(\frac{\Nre}{2}(1+3\wre)(\delta+n_{w}-2)\right)\right)^{\frac{1}{(6+\delta-4\nu_2-n_w)}} 
 \end{align}
where ~\[\mathcal{C}= \left(\frac{M_{\rm pl}}{\mathcal{A}_3\,\He}\right)^2\frac{15(3-\nu_2)(4\nu_2-5)}{8(1+2\nu_2)}(1+3\wre)^{(2-\nw)}\Gamma^2\l(\frac{5+3\wre}{2+6\wre}\r)\left(\frac{\Gamma^2(l+ \frac{\delta}{2})\Gamma^2\l( \frac{\delta}{2}\r)}{2^{1-2l-2\delta}\Gamma^2(1-l)\Gamma^2(l)}\right)\]
For example, $\wre=3/5$ and the coupling strength $\xi=2.9$, the primary GW spectrum becomes dominant for  $\mbox{f}_{\rm SB}=k_{\rm SB}/2\pi\sim 1.84\times 10^{-3}$ Hz, for $\wre=2/3$ and $\xi= 2.6$, this occurs around $\mbox{f}_{\rm SB}\sim 7.17\times 10^{-2}$ Hz, for $\wre=5/7$ and $\xi=2.5$, this occurs around $\mbox{f}_{\rm SB}\sim 3.43$ Hz, and for   $\wre=9/11$ and $\xi=2.4$, we get $\mbox{f}_{\rm SB}\sim 1.52\times 10^3$ Hz. All the features can be observed in the figure (\ref{gwspectrumfig}).  Further, in all the plots, the blue-tilted nature in the low-frequency regime($k<\kre$), and the mixed nature(red-tilted for $k_{\rm re}<k< k_{\rm SB}$ and blue-tilted for $ k_{\rm SB}<k<\ke$) in the intermediate frequency regime($\kre<k<\ke$) nicely follow the spectral nature as given in Eq.(\ref{break}) for all EoS. It is important to note that in the range $k_{\rm SB}<k<\ke$, the PGW spectral index $\nw$ will remain valid up to the maximum frequency $k_{\rm end}$. However, if one assumes modes $k>k_{\rm end}$, that remain sub-horizon during inflation, the spectral density can be assumed to be convergent. For example, in  
\cite{Chakraborty:2025zgx}, 
we have shown that for massless minimally coupled scalar field %
spectral behavior of the number density assumes convergent form $|\beta_k|^2\propto k^{-6}$, irrespective of the EoS. 
Furthermore, the behavior of GW is similar to that of a massless minimally coupled scalar field; hence, the PGW spectrum will also remain convergent beyond $k>\ke$. Thus 
it does not violate the $\Delta N_{\rm eff}$ bound at any higher frequency.
Interestingly, we also observe that the enhanced GW spectra in the intermediate frequency range pass through various sensitivity curves like LISA, BBO, DECIGO, and some parts of ET also. This detection prospect indeed opens up a possibility to probe such a reheating scenario through future GW observatories.  

\begin{figure}[t]
     \begin{center}
\includegraphics[scale=0.3]{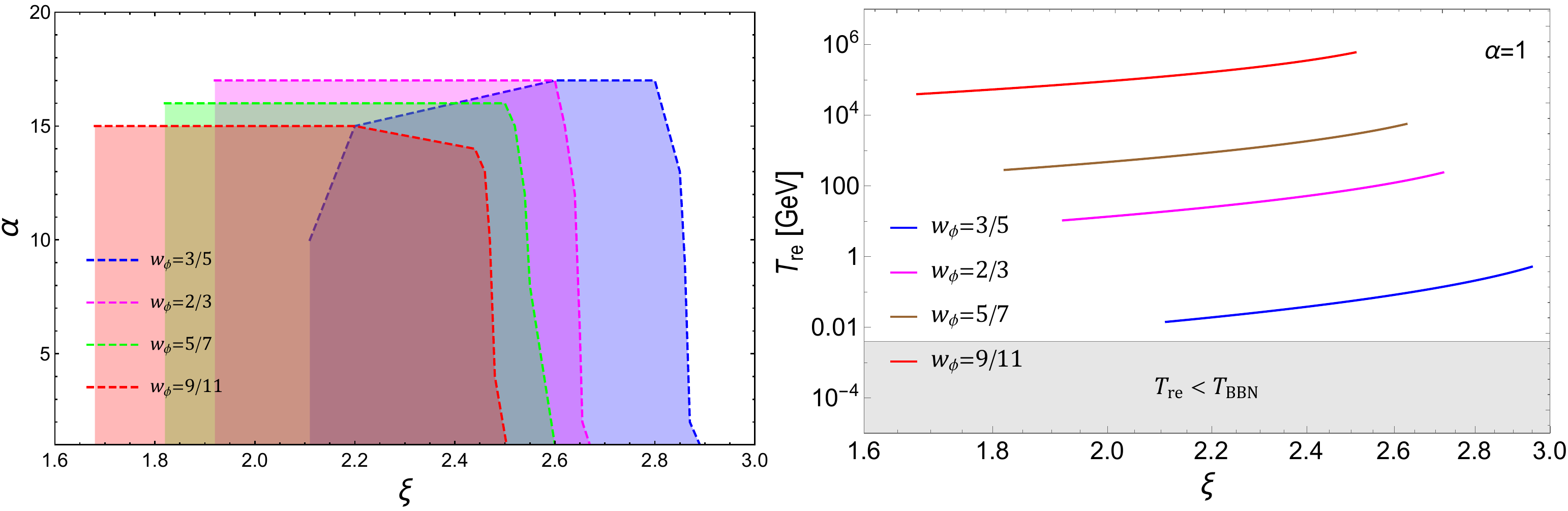}
\caption{\textit{\textbf{Left panel}: Figure represents the allowed $\xi$ vs $\alpha$ parameter space in this infrared gravitational reheating dynamics for different reheating EoS. In this figure, the color-shaded regions provide an admissible parameter set ($\xi, \alpha$)  for different EoS based on the bounds, $r_{0.05}\lesssim 0.038$,~ $\mathcal{P}_{\mathcal{S}}(k_{\ast})\lesssim 8.3\times 10^{-11}$,~$\Omega_{\rm gw}h^2\lesssim 9.54\times 10^{-7}$, and~$T_{\rm re}\gtrsim 4$ MeV. \textbf{Right panel:} Figure represents the variation of the infrared gravitational reheating temperature with the non-minimal coupling strength within its allowed range($\xi_{\rm min}\leq \xi\leq\xi_{\rm max}$) for different reheating EoS. }} 
\label{reheatingparameterspacefig}
\end{center}
\end{figure}
\subsection{Constraining \enquote{$\xi$} vs \enquote{$\alpha$} parameter space based on observational bounds :}

\begin{figure}[t]
     \begin{center}
\includegraphics[scale=0.29]{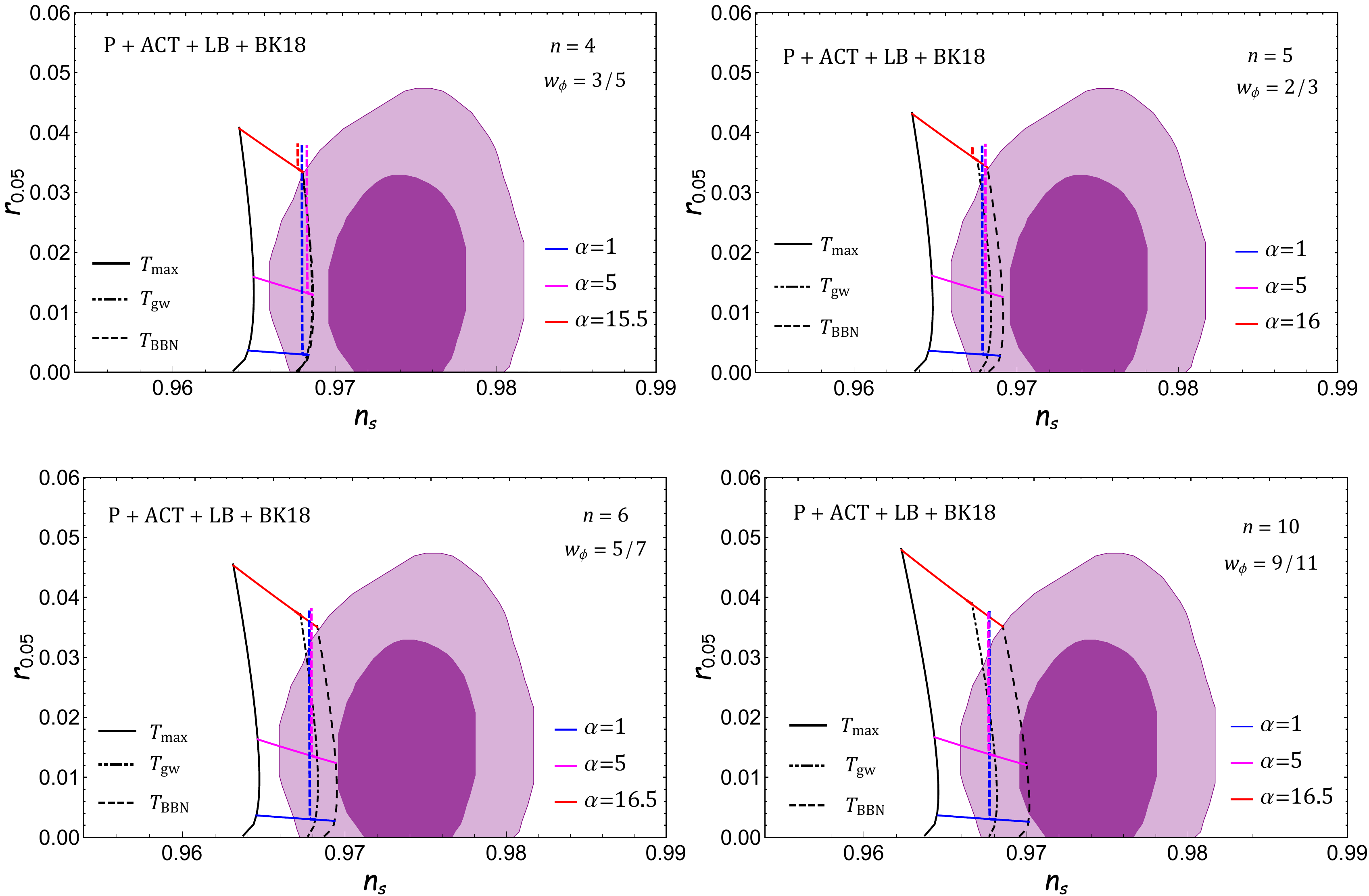}
\caption{\textit{ Figure represents the $n_s, r_{0.05}$ parameters predicted by the $\alpha$-attractor E-model (solid blue, magenta, red lines), and the infrared gravitational reheating dynamics(dashed blue, magenta, red lines) within its allowed coupling regime($\xi_{\rm min}\leq\xi\leq\xi_{\rm max}$) in the $n_{s}-r_{0.05}$ contour based on the 
latest Planck, ACT, DESI, and
 BICEP/Keck combined (\textbf{P+ACT+LB+BK18})(\textcolor{purple}{Purple}) data sets for four different reheating EoS. Here $T_{\rm max} \sim 10^{15}$ GeV (solid black line), $T_{\rm BBN}=4$ MeV (dashed black line), and $T_{\rm gw}$(black dot-dashed line) is the minimum reheating temperature satisfying the $\Delta N_{\rm eff}$ bound of PGW.
 }}
\label{fignsrinfraredACT}
\end{center}
\end{figure}

Finally, we constrain the theory parameter space $(\xi,\alpha)$ appropriately taking into account reheating dynamics and the latest ACT, DESI data as depicted in Fig.(\ref{reheatingparameterspacefig}). The figure indicates that for a given EoS, there exists a maximum limit of $\alpha$ associated with each value of $\xi$ within the allowed range of coupling strength, $\xi\leq \xi_{\rm max}$. For smaller $\xi$ values 
primary GW (see Eq.(\ref{rprimary}),\,(\ref{eq:gws_pri})) dominates over the secondary one, and hence the upper limit on $\alpha$ is nearly constant 
for any EoS. Such behavior can be clearly observed in the left panel of Fig.(\ref{reheatingparameterspacefig}). For example, we obtain $\alpha_{\rm max}\simeq 17$ for the coupling strength in the range $1.92\lesssim \xi\lesssim 2.6$ for $\wre=2/3$. Upon increasing the coupling beyond $\xi\gtrsim 2.6$, secondary GW production is expected to become significant, causing the rapid growth of $r_{0.05}$, and also the enhancement of the large-scale field amplitude quickly violates the isocurvature bound, setting
a maximum value of $\xi\sim 2.7251$, and the parameter $\alpha$ sharply drops to unity at the maximum value $\xi_{\rm max}$. \textit{In the proximity of $\xi=\xi_{\rm max}$, a sudden and rapid rise of $r_{0.05}$ is a prominent signature of the production of secondary gravitational waves (see dashed color lines in Fig.(\ref{fignsrinfraredACT})). 
Any $\xi<1.92$ is discarded because of the overproduction of primary gravitational wave, violating the $\Delta N_{\rm eff}$ bound.} For $\wre=2/3$, we therefore get a very narrow allowed range of coupling parameter, $1.92\lesssim\xi\lesssim 2.7251$.

Likewise for the other two EoS $\wre=(5/7,~ 9/11)$, the allowed range of $\xi$ and the upper limit of $\alpha$ is ($1.82\lesssim \xi\lesssim 2.6256,~ 16$) and ($1.68\lesssim \xi\lesssim 2.5041,~ 15$) respectively. Unlike the above three EoS, for $\wre=3/5$, the constancy of the upper limit of $\alpha$ for smaller $\xi$ is absent owing to the satisfaction of the lowest reheating temperature bound $T_{\rm BBN}$. For a smaller $\xi$, maximum $\alpha$ is obtained taking the lowest temperature bound into account, which lowers the maximum possible $\alpha$ for coupling in the range $2.11\lesssim\xi\lesssim 2.6$ as shown in the blue line in the left panel of Fig.(\ref{reheatingparameterspacefig}). At $\xi=2.6$, $\alpha$ becomes maximum $\alpha_{\rm max}=17$, and further increase of $\xi$ gradually lowers $\alpha$ as discussed before. For this EoS, we obtain the allowed $\xi$ range and upper limit of $\alpha$ as ($2.11\lesssim \xi\lesssim 2.9483,~ 17$).\\ 
In Fig.(\ref{fignsrinfraredACT}) we summarize the final constraints on the model parameters taking all the relevant constraints into account, namely the bound on tensor-to-scalar ratio ($r_{0.05}\lesssim 0.038$), isocurvature ($\Ps(k_{\ast})\lesssim 8.3\times 10^{-11}$), BBN constraint on $\Delta N_{\rm eff} < 0.17$ ($\Omega_{\rm gw}h^2\lesssim 9.54\times 10^{-7}$), and the lowest possible reheating temperature $\Tre= T_{\rm BBN} \sim 4$ MeV), 
 in light of combined P+ACT+LB+BK18 observational data sets (see the references \cite{ACT:2025tim, ACT:2025fju}). 
 In Fig.(\ref{fignsrinfraredACT}), using Eq.(\ref{rprimary}), we have presented the $\alpha$-attractor E-model prediction of $n_s, r_{0.05}$ parameters for varying $\alpha$(solid color lines) on top of the observational contour provided by the latest \textit{Planck}, ACT, DESI, and
 BICEP/Keck combined (P+ACT+LB+BK18) observational data sets \cite{ACT:2025tim, ACT:2025fju}. These predictions span reheating temperatures from the maximum reheating temperature, $T_{\rm max} \sim 10^{15}$ GeV (solid black lines) to the BBN temperature, $T_{\rm BBN}=4$ MeV (dashed black lines). When $\wre \geq 3/5$, the minimum reheating temperature, $T_{\rm gw}$, derived from satisfying the BBN constraint of the maximum PGW amplitude at the scale $k = \ke$, is stronger than the BBN temperature\cite{Chakraborty:2024rgl}.
  This is illustrated in Fig.~(\ref{fignsrinfraredACT}) with a black dot-dashed line. Within the full parameter range, the model clearly falls within the $2\sigma$ region of the observational contour and outside of $1\sigma$ region. In this Fig.(\ref{fignsrinfraredACT}), with every solid color line for $\alpha$-attractor E-model, we associate a dashed color line representing the predictions of $n_s, r_{0.05}$ parameters of the infrared reheating dynamics. It is seen that for $\wre\geq 3/5$, each dashed color line meets the solid line where it cuts the $T_{\rm gw}$ black dot-dashed line. This feature is caused by setting the lower limit($\xi_{\rm min}$) of coupling strength respecting the $\Delta N_{\rm eff}$ bound of PGW as discussed earlier. With further increase of $\xi$, the infrared reheating prediction of $r_{0.05}$ parameters noticeably departs from the $\alpha$-attractor model prediction, showing a steep rise towards the maximum $r_{0.05}\simeq0.038$ at $\xi\simeq \xi_{\rm max}$ owing to the substantial growth of secondary tensor fluctuations as already stated. For $\wre= 5/7,\,9/11$ and $\alpha=16.5$, as the maximum contribution to $r_{0.05}$ comes from PGW production (Eq.(\ref{rprimary})), we find overlapping red dashed lines on the solid lines, as obvious in the bottom panel of Fig.(\ref{fignsrinfraredACT}).       
 The tight constraint represented by dashed color lines(blue, magenta, red) can be observed lying in the 2$\sigma$ region of the $n_s-r_{0.05}$ reported by the ACT collaboration. 
\paragraph{\underline{Final infrared reheating temperature($\Tre$) vs $\xi$ parameter space} :}
Subject to all relevant latest observational bounds, we have shown the variation of reheating temperature within the allowed range of coupling strength ($\xi_{\rm min}\leq\xi\leq\xi_{\rm max}$) for different $\wre$ in the right panel of Fig.(\ref{reheatingparameterspacefig}). In the present \textit{infrared gravitational reheating} scenario, we find the lowest possible EoS $\wre\simeq 3/5$, which is consistent with the bounds ($T_{\rm BBN}$,~ $\Delta N_{\rm eff}$,~ $r_{0.05}$,~$\Ps(k_{\ast})$). In the right panel of Fig.(\ref{reheatingparameterspacefig}), 
$\Omega_{\rm gw}h^2\lesssim 9.54\times 10^{-7}$ imposes the lower limit $\xi_{\rm min}$ for all four EoS. Whereas, 
for all the EoS,
$\xi_{\rm max}$ is set by the tensor-to-scalar ratio bound as mentioned earlier. This indeed is a striking improvement of the reheating parameter space compared to the non-perturbative minimal gravitational reheating scenario \cite{Chakraborty:2025zgx}. In the minimal scenario, although we achieved $T_{\rm BBN}$, we failed to satisfy the PGW $\Delta N_{\rm eff}$ constraint for $\wre\simeq 3/5$. Whereas in the non-minimal scenario, we respect all these relevant observational bounds in the entire range $\wre \gtrsim 3/5$ and achieve successful reheating and a notable induced GW signal.\\  

\section{Conclusion}\label{secconclusion}

Inflation is intimately tied to the physics of reheating. While constraining inflation via observation, one must take into account the physics of reheating. Over the years different reheating mechanisms have been proposed. 
In this work, we have proposed a new reheating scenario 
driven by non-minimally coupled ($\xi\chi^2R$) scalar field. 
Taking $\alpha-$attractor as a potential candidate of inflation model, and embedding our proposed reheating scenario we finally place constraint on the model parameters in light of the latest ACT, DESI results. 

 
\underline{a) Production of scalar fluctuations through infrared instability and reheating: }
Non-minimal coupling induces super-horizon instability in the scalar field both during and after inflation. Such instability produces large infrared fluctuations, and upon entering the post-inflationary phase, those modes successfully reheat the universe.
For $\wre<1/3$, we find reheating cannot be achieved due to the fact that inflaton field dilutes much slower than the infrared fluctuation.
On the contrary, for $\wre > 1/3$, we find infrared fluctuation can successfully reheat the universe due to faster dilution of inflaton field. 
We have compared our non-perturbative infrared results with that of the perturbative one. In the perturbative reheating, one needs to calculate the width of the inflaton decaying into radiation. In the present context, since radiation is non-minimally coupled, decay widths are computed both in Jordan and Einstein frames. 
Important to note that perturbative reheating describes the dynamics of sub-Hubble modes and always ignores the modes that are outside the horizon. Therefore, while studying the post-inflationary reheating phase, one must take into account the super-Hubble modes separately, which are usually ignored in the literature. And in the present context, we indeed found that it is the infrared modes that actually dominate over the perturbative sub-Hubble modes above a critical coupling strength. 

This essentially results in a noticeable difference in the $\Tre$ vs $\xi$ parameter space as shown in Fig.(\ref{comparisonfig}) as compared to the perturbative results.
In Fig.(\ref{comparisonfigJorEin}), we have also shown a comparison between the Jordan and Einstein frame reheating parameter space. Interestingly, near the conformal limit, we find a discrepancy in the predictions of these two frames, which justifies the non-equivalence between these two frames in the non-minimal coupling induced reheating scenario.\\ In this context, we would like to mention one important point. In our present model, the parametric resonance can be sourced by two factors: i) Oscillation of the curvature scalar $R$ along with sufficiently high non-minimal coupling strength $\xi>50$, that can lead to strong non-perturbative resonance particularly for the modes $k\sim \ke$, as discussed in \cite{Garcia:2023qab}, and ii) For equation of state $\wre > 1/3$, there exists self resonance phenomena\cite{Lozanov:2017hjm}
that can give rise to radiation domination early. This self-resonance can have an impact depending upon its strength, which needs detailed investigation. We can compare the efficiency of the self-resonance dynamics with the present reheating scenario, which is left for our future endeavour.

\underline{b) Impact of ACT, DESI on the $\alpha-$attractor along with infrared reheating 
: }
Besides being a dominant radiation component to reheat the universe, this massless scalar also generates significant anisotropies to source notable induced gravitational waves. We find that for $\wre>1/3$, the growing $\xi$ of the field causes larger tensor and isocurvature fluctuations at the CMB scale. Therefore, the latest  (P+ACT+LB+BK18) bound  $r_{0.05}\lesssim 0.038$, and the \textit{Planck} 2018 bound $\Ps(k_{\ast})\lesssim 8.3\times 10^{-11}$ put tight constraint on maximum $\xi$. Furthermore, for a given EoS, low $\xi$ causes low reheating temperature, and hence enhances the duration of reheating, producing stronger high-frequency gravitational waves. Hence, the BBN bound $\Omega_{\rm gw}h^2\lesssim 9.54\times 10^{-7}$ imposes further constraint on minimum value of $\xi$. Thus, different observations tighten the associated reheating dynamics. We find the lowest possible EoS, $\wre=3/5$, which successfully reheats the universe satisfying all relevant bounds ($T_{\rm BBN},~ r_{0.05},~\Ps(k_{\ast}),~\Delta N_{\rm eff}$) in the coupling range $2.11\lesssim \xi\lesssim 2.9483$. After considering all the latest bounds, 
for a given EoS, $\wre\geq 3/5$, we find that the small-scales (UV) analyzed by perturbative treatment give higher reheating temperature than the IR modes in a larger portion within the allowed regime($\xi_{\rm min}\leq\xi\leq\xi_{\rm max}$) of the reheating parameter space.\\

\underline{c) Detectability of the GW signal:}
Furthermore, we have found a distinctive gravitational wave spectrum for different reheating parameters($\wre,~\alpha,~\Tre,~\xi$) which is strong enough to be detected by various GW observatories(see Fig.(\ref{gwspectrumfig})), allowing for more robust constraints on the coupling parameters and the reheating dynamics in the near future GW experiments. 


\section{Acknowledgments}
AC and RM gratefully acknowledge the Ministry of Human Resource Development, Government of India (GoI), for financial assistance. 
DM wishes to acknowledge support from the Science and Engineering Research Board~(SERB), Department of Science and Technology~(DST), Government of India~(GoI), through the Core Research Grant CRG/2020/003664. 

\appendix
\section{Calculation of the perturbative production rate :}\label{appenproductionrate}
In the computation of perturbative production rates for both minimal and non-minimal cases, we follow the metric signature $(+,-, -, -,)$ for computational simplicity. The final results will certainly remain independent of the metric signature.
\subsection{For minimal case}
\begin{figure}
    \centering    \includegraphics[width=\linewidth]{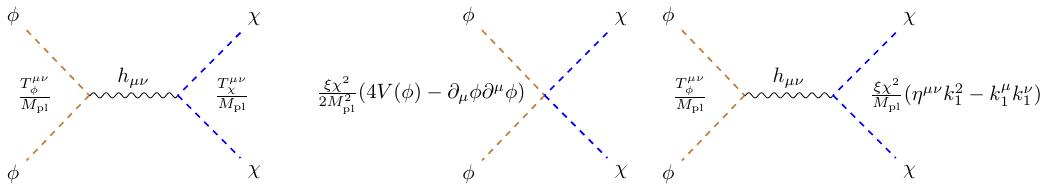}
    \caption{Feynman diagrams illustrating the perturbative gravitational production of the scalar field ($\chi$) in the minimal scenario (left), the Einstein frame (middle), and the Jordan frame (right) in the non-minimal coupling scenarios.}
\label{fnd}
\end{figure}

The energy-momentum tensor for the minimally coupled scalar inflaton field is given by,
\begin{equation}\label{Tmunuphi}
T_{\mu\nu}^{\phi} =  \partial_\mu \phi \partial_\nu \phi - \frac{1}{2} \eta_{\mu\nu} (\partial^\alpha \phi \partial_\alpha \phi - V(\phi)).
\end{equation}
For the minimally coupled scalar fluctuation, the energy-momentum tensor is given by
\begin{equation}\label{Tmunuchi}
T_{\mu\nu}^{\chi} = \partial_\mu \chi\partial_\nu \chi - \frac{1}{2} \eta_{\mu\nu} (\partial^\alpha \chi \partial_\alpha \chi - m_\chi^2\,\chi^2).
\end{equation}
Note that minimal coupling with gravity always leads to a three-point vertex at the tree level(see the left panel of Fig.(\ref{fnd})). For the minimal case, the gravitational interaction takes the following form.
\begin{equation}\label{gravinteractionmin}
    \mathcal{L}_{\rm int}^{}= -\frac{ h^{\mu\nu}}{M_{\rm pl}}\left(T^{\phi}_{\mu\nu}+T^{\chi}_{\mu\nu}\right) 
\end{equation}
where $T^{\phi}_{\mu\nu}$, and $T^{\chi}_{\mu\nu}$ are given by Eqs.(\ref{Tmunuphi}) and (\ref{Tmunuchi}).
Further treating the inflaton as a classical background, the transition probability turns out to be proportional to the energy-momentum tensor as 
\begin{equation}{\label{mphi}}
\mathcal M_{\mu\nu}^{\phi} = -i\frac{T_{\mu\nu}^{\phi}}{M_{\rm pl}} = -\frac{i}{M_{\rm pl}} \left\{ \partial_{\mu} \phi\, \partial_{\nu} \phi - \frac{1}{2} \eta_{\mu\nu} \left( \partial^{\alpha} \phi \,\partial_{\alpha} \phi - V(\phi) \right) \right\}
\end{equation}
for $\phi\,\phi\,h_{\mu\nu}$ vertices (see left panel of Fig.~\ref{fnd}). 
For $\chi^2\,h_{\mu\nu}$ vertices, however, it is,
\begin{equation}
\mathcal{M}_{\rho\sigma}^{\chi} = - \frac{2\,i}{ M_{\rm pl}} \left( p_{1\rho} p_{2\sigma} + p_{1\sigma} p_{2\rho} - \eta_{\rho\sigma} (p_1 \cdot p_2 + m_\chi^2) \right),
\end{equation}
where \enquote{$p_1$} and \enquote{$p_2$} are the final state $\chi$ particle's four-momenta. The final scattering amplitude is expressed as,
\begin{equation}
\mathcal{M}^{\phi\chi} = \mathcal{M}_{\mu\nu}^{\phi}\, \Pi^{\mu\nu\rho\sigma}\, \mathcal{M}_{\rho\sigma}^{\chi},
\end{equation}
where $\Pi^{\mu\nu\rho\sigma} $ is the graviton propagator for the canonical field $h_{\rm\mu\nu}$ with momentum $\sqrt{s}$,
\begin{equation}
\Pi^{\mu\nu\rho\sigma} = \frac{1}{2\, s} \left( \eta^{\mu\rho} \eta^{\nu\sigma} + \eta^{\mu\sigma} \eta^{\nu\rho} - \eta^{\mu\nu} \eta^{\rho\sigma} \right).
\end{equation}
where \enquote{$s$} is the Mandelstam variable defined in terms of the background condensate energy. Utilizing the above expression of the propagator we obtained,
\begin{eqnarray}
&& M^{\phi}_{\mu\nu}\, \Pi^{\mu\nu\rho\sigma} = \frac{1}{M_{\rm pl}\, s} \left[ \partial^{\rho} \phi \partial^{\sigma} \phi - \eta^{\rho\sigma} V(\phi) \right], \\
&& \mathcal{M}^{\phi\chi} = \frac{2}{M_{\rm pl}^2 s} \left( \partial^\rho \phi \partial^\sigma \phi - \eta^{\rho\sigma}  V(\phi) \right) \left( p_{1\rho} p_{2\sigma} + p_{1\sigma} p_{2\rho} - \eta_{\rho\sigma} (p_1\cdot p_2 + m_\chi^2) \right).
\end{eqnarray}
Expanding and simplifying the above equation, we have
\begin{equation}
\mathcal{M}^{\phi\chi} = \frac{2\,i}{ M_{\rm pl}^2\,s} \left[ (\partial^\rho \phi \, p_{1\rho}) (\partial^\sigma \phi \, p_{2\sigma}) + (\partial^\rho \phi \, p_{2\rho}) (\partial^\sigma \phi \, p_{1\sigma}) - 2 p_1\cdot p_2  V(\phi) + 4  V(\phi) (p_1\cdot p_2 + m_\chi^2)\right].
\end{equation}
For a homogeneous field \( \phi(t) \), \( \partial_\mu \phi = \dot{\phi}\) and spatial derivatives vanish. Substituting this,
\begin{equation}\label{m1}
\mathcal{M}^{\phi\chi} = \frac{2\,i}{ M_{\rm pl}^2\,s} \left[2\, \dot{\phi}^2\, p_1^0\, p_2^0 - \dot{\phi}^2 (p_1\cdot p_2 + m_\chi^2) + 2 p_1\cdot p_2\, V(\phi) + 4\,m_\chi^2\, V(\phi) \right].
\end{equation}
For the inflaton condensate, we can use the transition amplitude \(\mathcal{M}_\nu\) for each oscillating field mode of \(\phi\), defined the main text. In this case, the four-momentum of the \(\nu\)-th oscillation mode is given by $
p_\nu = \sqrt{s} = (E_\nu, 0)$
where \(E_\nu=\nu\,\omega\) represents the energy of the \(\nu\)-th oscillation mode. The four-momenta \( p_1 = (E_\nu/2, \vec{p}) \) and \( p_2 = (E_\nu/2, -\vec{p}) \), \( s = (p_1 + p_2)^2 = E_\nu^2 \) and $p_1\cdot p_2=E_\nu^2/2-m_\chi^2$. Finally from Eq.(\ref{m1}), we obtain
\begin{equation}
|\mathcal{M}^{\phi\chi}_\nu|^2 = \frac{1}{ M_{\rm pl}^4}  V(\phi)^2 \left( 1 + \frac{2 m_\chi^2}{E^2_\nu} \right)^2.
\end{equation}
Using $V(\phi)=V(\phi_0)\,\mathcal{P}^{2n}(t)=\rho_\phi\sum_{\nu}\mathcal{P}_\nu^{2n}\,e^{-i\nu\omega t}$ and $\rho_\phi\simeq V(\phi_0)$,
\begin{equation}{\label{mvf}}
 |\mathcal{M}^{\phi\chi}_\nu|^2 = \frac{\rho^2_\phi}{ M_{\rm pl}^4}\left( 1 + \frac{2\, m_\chi^2}{\nu^2\,\omega^2} \right)^2 \lvert\mathcal P^{ 2n}_\nu\rvert^2.\end{equation}
 To calculate the decay width, we have used the following relation \cite{PhysRevD.101.123507},
\begin{equation}\label{eq:decayrate}
      \Gamma_\phi = \frac{1}{8\pi (1 + w_\phi ) \rho_\phi}\sum_{\nu}E_\nu\,\lvert\mathcal{M}^{\phi\chi}_\nu\rvert^2\left(1-\frac{4\,m_\chi^2}{E^2_\nu}\right)^{1/2}
 \end{equation}
and the final expression for the decay width will be 
\begin{equation}
      \Gamma_\phi = \frac{\rho_\phi}{8\,\pi (1 + w_\phi)\mpl^4}  \sum_{\nu}\nu \,\omega \,|\mathcal P_\nu^{2n}|^2  \left(1 - \frac{4 m_{\chi}^2}{\nu^2 \omega^2}\right)^{\frac{1}{2}} \left(1 + \frac{2 m_\chi^2}{\nu^2 \omega^2}\right) .
 \end{equation}
 For massless radiation $m_\chi=0$ and $\omega=\gamma\,m_\phi(t)$, so the production rate
\begin{equation}\label{eq:productionrate}
    \boxed{Q(t)=(1+\wphi)\,\Gamma_\phi\,\rphi=\frac{\rho^2_\phi\,m_\phi}{8\pi\mpl^4}  \sum_\nu\gamma \,\nu \,|\mathcal P_\nu^{2n}|^2}
\end{equation}

\subsection{For non-minimal case}
\subsubsection{\underline{Einstein frame analysis }:}
In the presence of non-minimal coupling between the massless scalar and gravity, we write down the total action of the inflaton-scalar field system in the Jordan frame as follows:
\begin{equation}\label{Jordanaction}
    S=\int \sqrt{-g}\,d^4x\left(-\frac{M_{\rm pl}^2}{2}\Omega^2 R_{}+\frac{1}{2}\partial_{\mu}\phi\partial^{\mu}\phi-V(\phi)+\frac{1}{2}\partial_{\mu}\chi\partial^{\mu}\chi\right),
\end{equation}
with the conformal factor $\Omega^2$ is given by
\begin{equation}\label{conformalfactor}
  \Omega^2\equiv \left(1+\frac{\xi \chi^2}{M_{\rm pl}^2}\right) .
\end{equation}
To deal with a non-minimally coupled system, It is convenient to switch to the Einstein frame by the following metric redefinition:
\begin{equation}\label{einsteinmetric}
  g_{\mu\nu}^{\rm E}=\Omega^2g_{\mu\nu}^{}  
\end{equation}
Under this transformation the Jordan frame Ricci scalar is transformed as \cite{Fujii:2003pa}
\begin{equation}\label{Einsteinricci}
   R=\Omega^2\left(R^{\rm E}+6\nabla_{\mu}\nabla^{\mu}{\rm ln}\Omega-6g^{\rm E \mu\nu}\partial_{\mu}{\rm ln}\Omega~\partial_{\nu}{\rm ln}\Omega\right).
\end{equation}
Second term in the above equation being the total divergence term will not contribute to the Einstein frame action. Therefore, eliminating the term we find the Einstein frame action as follows:
\begin{equation}\label{einsteinaction1}
  S_{\rm E}=\int \sqrt{-g^{\rm E}}d^4x\left(-\frac{M_{\rm pl}^2}{2}R^{\rm E}+\frac{1}{2 \Omega^2}\partial_{\mu}\phi\partial^{\mu}\phi+\frac{1}{2 \Omega^2}\partial_{\mu}\chi\partial^{\mu}\chi+\frac{6 M_{\rm pl}^2}{2}g^{\rm E \mu\nu}\partial_{\mu}{\rm ln}\Omega~\partial_{\nu}{\rm ln}\Omega-\frac{V(\phi)}{\Omega^4}\right)  
\end{equation}
As we are interested in the small-field limit $\frac{\xi~ \chi^2}{M_{\rm pl}^2}<<1$, the conformal factor(\ref{conformalfactor}) is approximated as
\begin{equation}\label{approxconformalfactor}
 \Omega\approx \left(1+\frac{\xi\chi^2}{2M_{\rm pl}^2}\right)  .
\end{equation}
Let us now simplify the Einstein frame inflaton-scalar field Lagrangian (\ref{einsteinaction1}) as follows:
\begin{align}\label{einsteinlagrangian}
  \mathcal{L}_{\rm E}&=\left(\frac{1}{2 \Omega^2}\partial_{\mu}\phi\partial^{\mu}\phi+\frac{1}{2 \Omega^2}\partial_{\mu}\chi\partial^{\mu}\chi+3 M_{\rm pl}^2~g^{\rm E \mu\nu}\partial_{\mu}{\rm ln}\Omega~\partial_{\nu}{\rm ln}\Omega-\frac{V(\phi)}{\Omega^4}\right)\nno\\
  &=\frac{1}{2 \Omega^2}\partial_{\mu}\phi\partial^{\mu}\phi+\frac{1}{2 \Omega^2}\partial_{\mu}\chi\partial^{\mu}\chi+\frac{3 M_{\rm pl}^2~g^{\rm E \mu\nu}}{\Omega^2}\left(\frac{\partial \Omega}{\partial \chi}\right)^2\partial_{\mu}\chi\partial_{\nu}\chi-\frac{V(\phi)}{\Omega^4}\nno\\
  &\approx \left(1-\frac{\xi\chi^2}{M_{\rm pl}^2}\right)\frac{\partial_{\mu}\phi\partial^{\mu}\phi} {2}+\left(\frac{3~\xi^2\chi^2}{M_{\rm pl}^2}+\frac{1}{2}\right)\left(1-\frac{\xi\,\chi^2}{M_{\rm pl}^2}\right)\partial_{\mu}\chi\partial^{\mu}\chi-V(\phi)\left(1-\frac{2~\xi\chi^2}{M_{\rm pl}^2}\right)\nno\\
  & \approx \left(\underbrace{\frac{1}{2}\partial_{\mu}\phi\partial^{\mu}\phi-V(\phi)+\frac{1}{2}\partial_{\mu}\chi\partial^{\mu}\chi}_{\text {free Lagrangian}}+\underbrace{\frac{\chi^2}{2}\partial_{\mu}\chi\partial^{\mu}\chi\left(\frac{6 \xi^2}{M_{\rm pl}^2}-\frac{\xi }{M_{\rm pl}^2}\right)}_{\text {non-canonical kinetic terms}}\right)+\underbrace{\frac{\xi\,\chi^2}{M_{\rm pl}^2}\left(2\,V(\phi)-\frac{1}{2}\partial_{\mu}\phi\,\partial^{\mu}\phi\right)}_{\text{effective interaction Lagrangian}~ \mathcal{L}_{\rm E}^{\rm eff}}+\text{h.o}
\end{align}
In the present study, as we are considering a massless scalar field without any self-interaction, the free Lagrangian part of $\chi$ field doesn't contain any potential part.  Evidently, in the Einstein frame Lagrangian (\ref{einsteinlagrangian}) we encounter the non-canonical kinetic terms of the massless scalar field $\chi$, which vanishes at $\xi=1/6$. The following effective interaction Lagrangian $\mathcal{L}_{\rm E}^{\rm eff}$ gives the production of massless scalars from the background inflaton condensate, $\ket{\phi}\rightarrow\ket{\chi\chi}$, which is non-vanishing at the conformal limit.
Working in the weak field limit, with an expansion in powers of the gravitational coupling $G$ and taking the terms up to the linear order of perturbation, we write the Einstein frame metric as follows \cite{Holstein:2006bh}:
\begin{align}\label{einsteinmetric1}
   g_{\mu\nu}^{\rm E}&=\eta_{\mu\nu}+2\frac{h_{\mu\nu}}{M_{\rm pl}}\nno\\
   g^{\rm E\,\mu\nu}&=\eta^{\mu\nu}-2\frac{h^{\mu\nu}}{M_{\rm pl}}\nno\\
   \sqrt{-g^{\rm E}}&=1+\frac{\eta^{\mu\nu}h_{\mu\nu}}{M_{\rm pl}}
\end{align}
where $\eta_{\mu\nu}$ is the flat-space metric, and $h_{\mu\nu}$ is the perturbation (graviton field) on top of the flat metric. Using the expansions of (\ref{einsteinmetric1}) in the action 
(\ref{einsteinaction1}) and from the free Lagrangian and non-canonical kinetic terms involving parts of the Einstein frame Lagrangian $\mathcal{L}_{\rm E}$, we obtain the following leading-order gravitational interactions.
\begin{equation}\label{gravinteraction1}
    \mathcal{L}_{\rm E}^{\rm int}\supset -\frac{ h_{\mu\nu}}{M_{\rm pl}}\left(T^{\phi\,\mu\nu}+T^{\chi\,\mu\nu}\left(1+\frac{6 \xi^2\,\chi^2}{M_{\rm pl}^2}-\frac{\xi \chi^2}{M_{\rm pl}^2}\right)\right) 
\end{equation}
where $T^{\phi}_{\mu\nu}$, and $T^{\chi}_{\mu\nu}$ are given by Eqs.(\ref{Tmunuphi}) and (\ref{Tmunuchi}). Setting $\xi=0$ in (\ref{gravinteraction1}), we can get back the minimal interaction (\ref{gravinteractionmin}) as expected. From the effective interaction Lagrangian $\mathcal{L}_{\rm E}^{\rm eff}$ we obtain the following gravitational interaction Lagrangian.
\begin{align}\label{gravinteraction2}
  \mathcal{L}_{\rm E}^{\rm int}\supset \left(\frac{\xi\,\chi^2}{M_{\rm pl}^2}\right)\left(2V(\phi)-\frac{1}{2}\partial_{\mu}\phi\partial^{\mu}\phi\right)+\left(\frac{\xi\,\chi^2}{M_{\rm pl}^2}\right)\left(\frac{h_{\mu\nu}}{M_{\rm pl}}\right)\left(\partial^{\mu}\phi\partial^{\nu}\phi+\eta^{\mu\nu}\left(2 V(\phi)-\frac{\partial_{\alpha}\phi\partial^{\alpha}\phi}{2}\right)\right) 
\end{align}
Finally, combining (\ref{gravinteraction1}) and (\ref{gravinteraction2}), we write the total Einstein frame interaction Lagrangian as follows:
\begin{align}\label{gravinteraction3}
   \mathcal{L}_{\rm E}^{\rm int}= -h_{\mu\nu}&\left(\underbrace{\frac{T^{\phi\,\mu\nu}}{M_{\rm pl}}+\frac{T^{\chi\,\mu\nu}}{M_{\rm pl}}}_{\text{minimal 
   interaction}}-\underbrace{\frac{\xi\,\chi^2}{M_{\rm pl}^3}\left(\partial^{\mu}\phi\partial^{\nu}\phi+\eta^{\mu\nu}\left(2 V(\phi)-\frac{\partial_{\alpha}\phi\partial^{\alpha}\phi}{2}\right)\right)+\frac{T^{\chi\,\mu\nu}}{M_{\rm pl}^3}\Big(6 \xi^2\,\chi^2-\xi \chi^2\Big)}_{\text{non-minimal gravitational interaction with}~ M_{\rm pl}^{-3}~ \text{suppression}}\right)\nno\\
&+\underbrace{\left(\frac{\xi\,\chi^2}{M_{\rm pl}^2}\right)\left(2V(\phi)-\frac{1}{2}\partial_{\mu}\phi\partial^{\mu}\phi\right)}_{\text{non-minimal gravitational interaction with}~M_{\rm pl}^{-2}~\text{suppression}}
\end{align}
We finally proceed to study the non-minimal production of massless scalar fluctuations from the background inflaton condensate with the leading $1/M_{\rm pl}^2$ suppressed term in the interaction Lagrangian (\ref{gravinteraction3}). Therefore, the $1/M_{\rm pl}^2$ suppressed interaction leads to the process $\ket{\phi}\rightarrow\ket{\chi\chi}$ with a transition amplitude(like an 
interaction process, $\phi\phi\rightarrow\chi\chi$, where $\phi$ is an oscillating condensate)
\begin{align}\label{xiamplitude}
   \mathcal{M}_{\xi}^{\phi\,\chi}&=-\frac{i\xi}{M_{\rm pl}^2}\left(2 V(\phi)-\frac{1}{2}\partial_{\mu}\phi\partial^{\mu}\phi\right)\nno\\
  \mathcal{M}_{\xi}^{\phi\,\chi} &=-\frac{i\xi}{M_{\rm pl}^2}\sum_{\nu=1}^{\infty}\left(2\rho_{\phi}\mathcal{P}_{\nu}^{2n}+\frac{2n(2n-1)\gamma^2\nu^2\rho_{\phi}|\mathcal{P}_{\nu}|^2}{2}\right)\nno\\
  |\mathcal{M}_{\xi}^{\phi\,\chi}|^2&=\left(\frac{\xi \rho_{\phi}}{M_{\rm pl}^2}\right)^2\sum_{\nu=1}^{\infty}\bigg{|}\left(2\mathcal{P}_{\nu}^{2n}+\frac{2n(2n-1)\gamma^2\nu^2|\mathcal{P}_{\nu}|^2}{2}\right)\bigg{|}^2
\end{align}
To reach the final line of the Eq.(\ref{xiamplitude}) we have utilized the equations (\ref{eq:Hubble}), (\ref{eq:phi}), (\ref{fre}) along with the relations \cite{Chakraborty:2023lpr, Chakraborty:2025zgx}
\begin{equation}\label{eq:para}
   \dot\phi(t)\approx -i\nu\phi_0(t)\omega(t)\mathcal{P}(t)\, ,~V(\phi)=\rho_{\phi}(t)\mathcal{P}^{2n}(t)\, ,\,m_{\phi}(t)= \sqrt{6n(2n-1)}H(a)\left(\frac{M_{\rm pl}}{\phi_0}\right)
\end{equation}
Using the non-minimal transition amplitude (\ref{xiamplitude}) in the equations (\ref{eq:decayrate}) and (\ref{eq:productionrate}), we finally obtain the following expression of non-minimal coupling-induced production rate.
\begin{equation}\label{eq:non-minimalproduction}
    \boxed{Q(t)=\frac{\xi^2\rho^2_\phi\,m_\phi}{8\pi\mpl^4}\sum_{\nu=1}^{\infty}\nu\gamma\bigg{|}\left(2\mathcal{P}_{\nu}^{2n}+\frac{2n(2n-1)\gamma^2\nu^2|\mathcal{P}_{\nu}|^2}{2}\right)\bigg{|}^2}
\end{equation}

\subsubsection{\underline{Jordan frame analysis :}}

 Considering only the inflaton-scalar fluctuation part in the Lagrangian, from (\ref{Jordanaction}) we write the Jordan frame action as follows:
\begin{equation}\label{Jordanaction1}
   S=\int \sqrt{-g} d^4x\left(\frac{1}{2}\partial_{\mu}\phi\partial^{\mu}\phi-V(\phi)+\frac{1}{2}\partial_{\mu}\chi\partial^{\mu}\chi-\frac{1}{2}\xi\chi^2 R\right) 
\end{equation}
Likewise, Einstein frame gravitational field perturbation (see Eq.(\ref{einsteinmetric1})), we also quantize the gravitational field by taking the linear order metric perturbation in the Jordan frame. Under this metric perturbation, the curvature scalar up to the first order metric perturbation is written as follows \cite{Holstein:2006bh}:
\begin{equation}\label{Jordanricci}
   R^{(1)}=\frac{2}{M_{\rm pl}}\left(\eta_{\alpha\beta}\partial_{\mu}\partial^{\mu}h^{\alpha\beta}-\partial_{\mu}\partial_{\nu}h^{\mu\nu}\right)
\end{equation}
Substituting (\ref{Jordanricci}) to the action (\ref{Jordanaction1}), we write the Jordan frame action as a function of the perturbed metric field as follows: 
\begin{align}\label{Jordanaction2}
  S=&\int d^4x \left(1+\frac{\eta^{\mu\nu}h_{\mu\nu}}{M_{\rm pl}}\right)\left(\frac{\left(\eta^{\mu\nu}-\frac{2 h^{\mu\nu}}{M_{\rm pl}}\right)\left(\partial_{\mu}\phi\partial_{\nu}\phi+\partial_{\mu}\chi\partial_{\nu}\chi\right)}{2}-V(\phi)+\frac{\xi\chi^2}{M_{\rm pl}}\left(\partial_{\mu}\partial_{\nu}h^{\mu\nu}-\eta_{\alpha\beta}\partial_{\mu}\partial^{\mu}h^{\alpha\beta}\right)\right)
\end{align}
The above action (\ref{Jordanaction2}) leads to the following gravitational interaction in the Jordan frame:
\begin{align}\label{Jordanintlagrangian}
  \mathcal{L}_{\rm int}=-\underbrace{\frac{h^{\mu\nu}}{M_{\rm pl}}\left(T^{\phi}_{\mu\nu}+T^{\chi}_{\mu\nu}\right)}_{\text{minimal interaction}}+\underbrace{\left(\frac{\xi\,\chi^2}{M_{\rm pl}}\right) \left(\partial_{\mu}\partial_{\nu}h^{\mu\nu}-\eta_{\alpha\beta}\partial_{\mu}\partial^{\mu}h^{\alpha\beta}\right)}_{\text{non-minimal interaction}~ \mathcal{L}_{\rm int}^{\xi}}
\end{align}
Likewise the Einstein frame, we can also generate the minimal gravitational interaction setting $\xi=0$ in (\ref{Jordanintlagrangian}). 
To study the non-minimal production of $\chi$ particles from the background inflaton condensate, we consider the second-order scattering matrix term, $S^{(2)}_{fi}$, which gives the gravity-mediated non-minimal production processes. The non-minimal coupling associated relevant second-order matrix element is written as follows:
\begin{align}\label{Smat}
  S^{\xi\,(2)}_{fi}=&\frac{\xi}{2 M_{\rm pl}^2}\int d^4x\int d^4y\bra{f}T\Big{[}h^{\mu\nu}(x)T^{\phi}_{\mu\nu}(x)\chi^2(y)\left(\partial_{\mu}\partial_{\nu}h^{\mu\nu}(y)-\eta_{\alpha\beta}\partial_{\mu}\partial^{\mu}h^{\alpha\beta}(y)\right)\nno\\
  &+h^{\mu\nu}(y)T^{\phi}_{\mu\nu}(y)\chi^2(x)\left(\partial_{\mu}\partial_{\nu}h^{\mu\nu}(x)-\eta_{\alpha\beta}\partial_{\mu}\partial^{\mu}h^{\alpha\beta}(x)\right)\Big{]} \ket{i} 
\end{align}
where \enquote{$i$} and \enquote{$f$} stand for initial inflaton condensate, $\ket{\phi}$, and final $\chi$ particles state, $\ket{\chi\chi}$, respectively. To proceed to compute these matrix elements, we first quantize the graviton field($h_{\mu\nu}$), and the scalar field($\chi$).
\begin{align}\label{fieldquantization}
 \hat{\chi}(x)=&\int d\Pi_k\left(\hat{a}_{\vec{k}}e^{-ik_{\mu}x^{\mu}}+\text{h.c}\right)\nno\\
 \hat{h}^{\mu\nu}(x)=&\sum_{\lambda=++,--}\int d\Pi_k\left(\epsilon^{\lambda\,\mu\nu}(k)\hat{a}_{\vec{k}}^{\lambda}e^{-ik_{\mu}x^{\mu}}+\text{h.c}\right)
\end{align}
where $d\Pi_k=\frac{d^3 k}{(2\pi)^3\sqrt{2k^0}}$ is the phase-space factor. The polarization tensor for the spin-2 graviton field and the creation, annihilation operators of the scalar field satisfy the following relations: 
\begin{align}\label{polarizationprop}
  &\sum_{\lambda=++,--}  \epsilon^{\lambda\,\mu\nu}(k)\,(\epsilon^{\lambda\,\alpha\beta}(k))^{\ast}=\frac{1}{2}\left(\eta^{\mu\alpha}\eta^{\nu\beta}+\eta^{\mu\beta}\eta^{\nu\alpha}+\eta^{\mu\nu}\eta^{\alpha\beta}\right)=P^{\mu\nu\alpha\beta}\nno\\
  &\big{[}\hat{a}^{\lambda}_{\vec{p}},\hat{a}^{\lambda^{\prime}\,\dagger}_{\vec{k}}\big{]}=(2\pi)^3\delta^3(\vec{p}-\vec{k})\delta^{\lambda\lambda^{\prime}}
\end{align}
Using (\ref{fieldquantization}) in (\ref{Smat}) we compute the expression of the matrix element associated with the first term $h^{\mu\nu}(x)T^{\phi}_{\mu\nu}(x)\chi^2(y)\left(\partial_{\mu}\partial_{\nu}h^{\mu\nu}(y)-\eta_{\alpha\beta}\partial_{\mu}\partial^{\mu}h^{\alpha\beta}(y)\right)$ as follows:\\
\begin{align}\label{Smat1}
  S^{\xi\,(2)}_{\phi\rightarrow\chi\chi}\supset& \frac{\xi}{M_{\rm pl}^2}\sum_{\lambda\lambda^{\prime}}\int d^4x\int d^4y  \int d\Pi_{k_1}\int d\Pi_{k_2}\int d\Pi_{p}\int d\Pi_{q} T^{\phi}_{\mu\nu}(x)\sqrt{p_1^{0} p_2^{0}} \bra{0}\left(\epsilon^{\lambda\,\mu\nu}(k_1)\hat{a}_{\vec{k}_1}^{\lambda}e^{-ik_{1\mu}x^{\mu}}+\text{h.c}\right)\nno\\
  &\left(\epsilon^{\lambda^{\prime}\,\alpha\beta}(k_2)\hat{a}_{\vec{k}_2}^{\lambda^{\prime}}e^{-ik_{2\alpha}y^{\alpha}}+\text{h.c}\right)\Big(\eta_{\alpha\beta}(k_2)_{\mu}(k_2)^{\mu}-(k_2)_{\alpha}(k_2)_{\beta}\Big)\ket{0}\bra{0}\hat{a}_{\vec{p}_1}\hat{a}_{\vec{p}_2}\left(\hat{a}_{\vec{p}}e^{-ip_{\mu}y^{\mu}}+\text{h.c}\right)\nno\\
&\left(\hat{a}_{\vec{q}}e^{-iq_{\mu}y^{\mu}}+\text{h.c}\right)\ket{0}
\end{align}
where $p_1,~p_2$ are the final state four-momenta. The contraction of the graviton operators finally leads to the following propagator\cite{Holstein:2006bh, Ahmed:2022tfm}.
\begin{equation}\label{gravitonprop}
   D^{\mu\nu\alpha\beta}(x-y)=\lim_{\epsilon\rightarrow0^{+}}\int \frac{d^4 k_1}{(2\pi)^4}\frac{i}{k_1^2+i\epsilon} e^{-ik_1(x-y)}P^{\mu\nu\alpha\beta}
\end{equation}
It gives 
\begin{equation}\label{Tmunuprop}
   T^{\phi}_{\mu\nu}P^{\mu\nu\alpha\beta}=\left(\partial^{\alpha}\phi\partial^{\beta}\phi-\eta^{\alpha\beta}V(\phi)\right) 
\end{equation}
Using (\ref{gravitonprop}) and (\ref{Tmunuprop}) in (\ref{Smat1}), after some straightforward computations we reach
\begin{align}\label{Smat2}
S^{\xi\,(2)}_{\phi\rightarrow\chi\chi}\supset& \frac{\xi}{M_{\rm pl}^2}\int d^4 x\int e^{i(p_2+p_2)y}d^4 y\times \lim_{\epsilon\rightarrow 0^{+}}\int \frac{d^4 k_1}{(2\pi)^4}\frac{i e^{-i k_1(x-y)}}{k_1^2+i\epsilon}\left(\partial^{\alpha}\phi\partial^{\beta}\phi-\eta^{\alpha\beta}V(\phi)\right)\nno\\
&\times \Big(\eta_{\alpha\beta}(k_1)_{\mu}(k_1)^{\mu}-(k_1)_{\alpha}(k_1)_{\beta}\Big)\nno\\
=& \frac{i\xi}{M_{\rm pl}^2}\int d^4 x e^{-ik_1x}\int d^4 y\int \frac{d^4 k_1}{(2\pi)^4}\frac{e^{i(k_1+p_1+p_2)y}}{k_1^2}\nno\\
&\times\Big(\partial^{\alpha}\phi\partial_{\alpha}\phi(k_1)_{\beta}(k_1)^{\beta}-\partial^{\alpha}\phi(k_1)_{\alpha}\partial^{\beta}\phi(k_1)_{\beta}-3V(\phi)(k_1)_{\beta}(k_1)^{\beta}\Big)
\end{align}
For homogeneous inflaton background, i.e., $\phi=\phi(t)$, we write inflaton-scalar field system's kinematics as
\begin{equation}\label{kinematics}
    p_1\,.\,p_2=\frac{s}{2},~~ p_1^0=p_2^0=\frac{\sqrt{s}}{2},~~ s=(p_1+p_2)^2
\end{equation}
where \enquote{$s$} is the Mandelstam variable defined earlier in the case of minimal production.
\\ Using the equations (\ref{eq:phi}), (\ref{eq:para}), and (\ref{kinematics}) in (\ref{Smat2}) we write
\begin{align}\label{Smat3}
   S^{\xi\,(2)}_{\phi\rightarrow\chi\chi}\supset& -\frac{3i\xi\rho_{\phi}}{M_{\rm pl}^2}\sum_{\nu}\mathcal{P}_{\nu}^{2n}\int d^4 x e^{-i(\nu\omega-p_1^0-p_2^0)t+i(p_1+p_2)_jx^j}\nno\\
  =& -\frac{3i\xi\rho_{\phi}}{M_{\rm pl}^2}(2\pi)^4\sum_{\nu}\mathcal{P}_{\nu}^{2n}\delta(\nu\omega-p_1^0-p_2^0)\delta^3(\vec{p}_1+\vec{p}_2) 
\end{align}
We obtain the same result from the second element of the S-matrix (\ref{Smat}) also. Combining these two, we write the final expression of the non-minimal coupling associated second-order S-matrix element as follows:
\begin{equation}\label{Smat4}
 S^{\xi\,(2)}_{\phi\rightarrow\chi\chi}=-\frac{6i\xi\rho_{\phi}}{M_{\rm pl}^2}(2\pi)^4\sum_{\nu}\mathcal{P}_{\nu}^{2n}\delta(\nu\omega-p_1^0-p_2^0)\delta^3(\vec{p}_1+\vec{p}_2)    
\end{equation}
Using the minimal interaction in the Jordan frame interaction Lagrangian (\ref{Jordanintlagrangian}), we can compute the gravity-mediated minimal production of massless $\chi$ particles from the background inflaton condensate. We find the second-order S-matrix element for the minimal process as follows \cite{Ahmed:2022tfm} :
\begin{equation}\label{Smatminimal}
    S^{\xi=0\,(2)}_{\phi\rightarrow\chi\chi}=\frac{i\rho_{\phi}}{M_{\rm pl}^2}(2\pi)^4\sum_{\nu}\mathcal{P}_{\nu}^{2n}\delta(\nu\omega-p_1^0-p_2^0)\delta^3(\vec{p}_1+\vec{p}_2) 
\end{equation}
Therefore, we write the full expression of the S-matrix element describing the gravitational production of massless scalar particles from the inflaton condensate in the Jordan frame as follows:
\begin{equation}\label{Smatfinal}
     S^{(2)}_{\phi\rightarrow\chi\chi}=\left(S^{\xi=0\,(2)}_{\phi\rightarrow\chi\chi}+S^{\xi\,(2)}_{\phi\rightarrow\chi\chi}\right)=\frac{(1-6\xi)i\rho_{\phi}}{M_{\rm pl}^2}(2\pi)^4\sum_{\nu}\mathcal{P}_{\nu}^{2n}\delta(\nu\omega-p_1^0-p_2^0)\delta^3(\vec{p}_1+\vec{p}_2)
\end{equation}
From (\ref{Smatfinal}), we write the total transition amplitude of this gravitational production process as
\begin{equation}\label{JordanM}
  \big{|}\mathcal{M}^{\phi\chi}\big{|}^2 =\frac{(1-6\xi)^2\rho_{\phi}^2}{M_{\rm pl}^4}\sum_{\nu=1}^{\infty}|\mathcal{P}_{\nu}^{2n}|^2  
\end{equation}
\textbf{\underline{Using the Feynman rule}}: We can also directly derive this transition amplitude by using the Feynman rule. In the Jordan frame, the Feynman diagram for the production of scalar fluctuations $\chi$ from the background condensate in the presence of non-minimal coupling $\xi$ is shown in Fig.(\ref{fnd}) (see the right most plot). So, the expression of partial amplitudes for $\phi\,\phi\,h_{\mu\nu}$ vertices is the same as we defined in Eq.(\ref{mphi}). Here, the partial amplitude for $\chi^2\,h_{\mu\nu}$ vertices is different, which is
\begin{eqnarray}
\mathcal{M}_{\rho\sigma}^{\chi}= - \frac{2\,i\,\xi}{ M_{\rm pl}}(\eta_{\rho\sigma}\, k_{1\alpha}\, k^\alpha_{1}- k_{1\rho}\, k_{1\sigma})\,,
\end{eqnarray}
where $k_{1}=(p_1+p_2)$ momentum of the propagator. Therefore, the final scattering amplitude is
\begin{equation}
\begin{aligned}
\mathcal{M}^{\phi\chi}_{\xi}& = \mathcal{M}_{\mu\nu}^{\phi}\, \Pi^{\mu\nu\rho\sigma}\, \mathcal{M}_{\rho\sigma}^{\chi}\\
&=\frac{2\,i\,\xi}{M_{\rm pl}^2 s} \left( \partial^\rho \phi \partial^\sigma \phi - \eta^{\rho\sigma}  V(\phi) \right)\,(\eta_{\rho\sigma}\, k_{1\alpha}\, k^\alpha_{1}- k_{1\rho}\, k_{1\sigma})\\
&=\frac{2\,i\,\xi}{M_{\rm pl}^2 s} \left[(\dot\phi^2\,k^2_1-\dot\phi^2\,k_1^0\,k_1^0)-(V(\phi)\eta_{\rho\sigma}\,\eta^{\rho\sigma}\,k^2_1+V(\phi)k^2_1)\right]\\
&=-\frac{6\,i\,\xi}{M_{\rm pl}^2}V(\phi)~~~[\mbox{where}~~k_1^2=s\,,k^0_1=\sqrt{s},\,\eta_{\rho\sigma}\,\eta^{\rho\sigma}=4]\\
\end{aligned}
\end{equation}
The total scattering amplitude is 
\begin{equation}
\begin{aligned}
\mathcal{M}^{\phi\chi}&=\mathcal{M}^{\phi\chi}_{\xi=0}+\mathcal{M}^{\phi\chi}_{\xi}\\
&=\frac{i\,(1-6\,\xi)}{M_{\rm pl}^2}V(\phi)\\
\end{aligned}
\end{equation}
which is the same as we find in Eq.(\ref{JordanM}).\,
Using this total transition amplitude (\ref{JordanM}) in the equations (\ref{eq:decayrate}) and (\ref{eq:productionrate}), we finally obtain the following expression of the Jordan frame total gravitational production rate.
\begin{equation}\label{Jordanproductionrate}
   \boxed{Q(t)=\frac{(1-6\xi)^2\rho^2_\phi\,m_\phi}{8\pi\mpl^4}  \sum_{\nu=1}^{\infty}\gamma \,\nu \,|\mathcal P_\nu^{2n}|^2} 
\end{equation}
Clearly, unlike the Einstein frame (\ref{eq:non-minimalproduction}), this production rate vanishes at the conformal limit, $\xi=1/6$.
\section{Computation of Isocurvature power spectrum :}\label{appenA}
For stiff EoS $\wre>1/3$, the large-scale fluctuation strength becomes significant in the range $\xi>3/16$. Therefore, in the present reheating scenario, we shall confine ourselves to this particular regime while computing the isocurvature power spectrum amplitude.\\ For $\wre>1/3$ and $\xi>3/16$, with the knowledge of $\alpha_k, \beta_k$ in this regime, 
we obtain the following long-wavelength post-inflationary field solutions at a very late time $k\eta>>1$ \cite{Chakraborty:2024rgl}.

\begin{align}\label{Xfieldsol}
     X^{\rm long}_k(\eta)\approx & \frac{\Gamma(\nu_2)\text{exp}(-\pi \tilde{\nu}_1/2)}{4\sqrt{2k_{\rm end}}}\left(\frac{3\mu-2\nu_2}{\sqrt{(3\mu-1)}}H^{(1)}_{\nu_1}(\kbar)+\kbar\sqrt{3\mu-1}\Big(H^{(1)}_{\nu_1-1}(\kbar)-H^{(1)}_{\nu_1+1}(\kbar)\Big)\right)\nonumber\\
  &\times\frac{ \text{cos}(k\eta)}{\kbar^{\nu_2+1/2}}\left(\frac{2}{3\mu-1}\right)^{\nu_2}
\end{align}
The time-derivative of this solution can be approximated as
\begin{align}\label{Xfielddiff}
 (X_k^{\rm long})^{\prime}(\eta)\approx & \frac{\Gamma(\nu_2)\text{exp}(-\pi \tilde{\nu}_1/2)\sqrt{k_{\rm end}}}{2\sqrt{2}} \sqrt{3\mu-1}\left(\frac{2}{3\mu-1}\right)^{\nu_2}\frac{\left(\pi+i\text{cosh}(\pi\tilde{\nu_1})\Gamma(1-i\tilde{\nu_1})\Gamma(i\tilde{\nu_1})\right)}{\pi\Gamma(i\tilde{\nu_1})}\nn\\
 & \times \l(\frac{\bar{k}}{2}\r)^{i\tilde{\nu_1}}\l(\bar{k}\r)^{\l(1/2-\nu_2\r)}\text{cos}(k\eta)\nn\\
 \Rightarrow \l|(X_k^{\rm long})^{\prime}(\eta)\r|^2 \approx & 2 \mathcal{A}_3 \ke \l(\frac{k}{\ke}\r)^{1-2\nu_2}\text{cos}^2(k\eta)
\end{align}
Therefore, in the massless limit $m_{\chi}\approx 0$, the integrand $P_X(p,|\vec{p}-\vec{k}|)$ in Eq.(\ref{PX1}) can be written as
\begin{equation}\label{PX2}
 P_X(p,|\vec{p}-\vec{k})=4\mathcal{A}_3^2 \ke^2 \l(\frac{p}{\ke}\r)^{1-2\nu_2}\l(\frac{|\vec{p}-\vec{k}|}{\ke}\r)^{1-2\nu_2}\text{cos}^2(p\eta)~\text{cos}^2(|\vec{p}-\vec{k}|\eta)   
\end{equation}
The expression of this integrand (\ref{PX2}) is true for any $\xi>3/16$ for EoS $\wre>1/3$. In the entire range $\xi>3/16$, total energy-density of the system $\rho_{\chi}$ will be different below($3/16<\xi<\xi_{\rm cri}$) and above($\xi>\xi_{\rm cri}$) the critical coupling $\xi_{\rm  cri}$ (See Equations (\ref{comovingenergy3}) and (\ref{comovingenergy5})). This creates a difference in the amplitude of the isocurvature power spectrum (\ref{isocurvpow}). We first express the integral (\ref{isocurvpow}) in terms of the general energy-density function $\rho_{\chi}$ to compute the integration. Plugging the expression 
(\ref{PX2}) into the integral (\ref{isocurvpow}) we have



\begin{align}\label{isointegral1}
    \Ps(k)&=
    \frac{4\mathcal{A}_3^2\ke^5}{(2\pi)^4\rho_{\chi}^2a^8}\l(\frac{k}{\ke}\r)^{5-4\nu_2}\text{cos}^2(p\eta)~\text{cos}^2(|\vec{p}-\vec{k}|\eta)\int _{k_{\rm min}}^{k_{\rm max}}p^2 dp\int_{-1}^{1}d\gamma \l(\frac{p}{k}\r)^{(1-2\nu_2)}\l(\frac{|\vec{p}-\vec{k}|}{k}\r)^{(1-2\nu_2)}\nn\\
   & \approx  
  \frac{\mathcal{A}_3^2\ke^8}{(2\pi)^4\rho_{\chi}^2a^8} \l(\frac{k}{\ke}\r)^{8-4\nu_2}\int_{u_{\rm min}}^{u_{\rm max}}u^{(3-2\nu_2)} du\underbrace{\int_{-1}^{1}d\gamma ~\l(1+u^2-2u\gamma\r)^{(1/2-\nu_2)}}_{\text{angular integral}}
\end{align}
where \enquote{$\gamma$} is the angle between two momentum vectors $\vec{p}$ and $\vec{k}$, $\text{cos}(\gamma)=\hat{p}.\hat{k}$. The dimensionless quantity $u$ is defined to be $u=\l(p/k\r)$ where $u_{\rm min}=\l(k_{\rm min}/k\r)$ and $u_{\rm max}=\l(k_{\rm max}/k\r)$, where $k_{\rm min}$ is considered to be the present-day horizon scale(smaller than CMB scale), and $k_{\rm max}$ is considered to be $\ke$. We have to first perform the angular integral before going to the momentum part.\\

\underline{\textit{Angular integral }:}\\

The angular integral is calculated to be
\begin{align}\label{angularint}
   {\cal I}_1&= \int_{-1}^{1}d\gamma ~\l(1+u^2-2u\gamma\r)^{(1/2-\nu_2)} = \frac{1}{u(3-2\nu_2)}\big[\l(1+u\r)^{(3-2\nu_2)}-\l(1-u\r)^{(3-2\nu_2)}\big]
\end{align}
Substituting the angular integral (\ref{angularint})
back to (\ref{isointegral1}) we have
\begin{equation}\label{isointegral2}
    \Ps(k)=
    \frac{\mathcal{A}_3^2\ke^8}{(2\pi)^4(3-2\nu_2)\rho_{\chi}^2a^8}\l(\frac{k}{\ke}\r)^{8-4\nu_2}\underbrace{ \int_{u_{\rm min}}^{u_{\rm max}}du~ \frac{\big[\l(1+u\r)^{(3-2\nu_2)}-\l(1-u\r)^{(3-2\nu_2)}\big]}{u^{2(\nu_2-1)}}}_{\text{momentum integral}}
    \end{equation}
\underline{\textit{Simplification of momentum integral} :}\\

Evaluating the following momentum integral we get
\begin{align}\label{momentumint}
  {\cal I}_2&= \int_{u_{\rm min}}^{u_{\rm max}}du~ \frac{\big[\l(1+u\r)^{(3-2\nu_2)}-\l(1-u\r)^{(3-2\nu_2)}\big]}{u^{2(\nu_2-1)}}\nn\\
  & =\frac{u^{3-2\nu_2}}{(3-2\nu_2)}\bigg[{}_2F_1\Big((2\nu_2-3),(3-2\nu_2);(4-2\nu_2);-u\Big)-{}_2F_1\Big((2\nu_2-3),(3-2\nu_2);(4-2\nu_2);u\Big)\bigg]_{u_{\rm min}}^{u_{\rm max}}\nn\\
  &= I_2(u_{\rm max}) -I_2(u_{\rm min})
\end{align}
where ${}_2F_1(a,b;c;z)$ is a Gaussian or ordinary Hypergeometric function with four arguments. In the long-wavelength regime($k<<\ke$), we simplify the following expression $I_2$ for two limits separately. \\

\underline{\textit{For upper limit $u_{\rm max}$ :}}\\

For large $u$ limit, the expression (\ref{momentumint}) can be approximated as 
\begin{align}\label{momentumintlarge}
  I_2(u_{\rm max})\approx \bigg(\frac{(-1)^{(2\nu_2-3)}\big(-1+(-1)^{(3-2\nu_2)}\big)\Gamma(4\nu_2-6)\Gamma(4-2\nu_2)}{(3-2\nu_2)\Gamma(2\nu_2-3)}+\frac{u_{\rm max}^{(6-4\nu_2)}}{(6-4\nu_2)}\big(1+(-1)^{(4-2\nu_2)}\big)\bigg)  
\end{align}
\underline{\textit{For lower limit $u_{\rm min}$ :}}\\

For small $u$ limit, expression (\ref{momentumint}) can be approximated as
\begin{align}\label{momentumintsmall}
   I_2(u_{\rm min})\approx \frac{(3-2\nu_2)}{(2-\nu_2)}u_{\rm min}^{(4-2\nu_2)}
\end{align}
Using Equations (\ref{momentumintlarge}) and (\ref{momentumintsmall}) in (\ref{isointegral2}), we write the final form of the isocurvature power spectrum amplitude 
in terms of total energy-density $\rho_{\chi}$ as follows:  
\begin{align}\label{isofinalamp1}
\Ps(k)=
\frac{\mathcal{A}_3^2\ke^8}{(2\pi)^4(3-2\nu_2)\rho_{\chi}^2a^8}\Big(I_2(u_{\rm max})-I_2(u_{\rm min})\Big)\l(\frac{k}{\ke}\r)^{(8-4\nu_2)}
\end{align}
\section{Comparing the perturbative approach in Jordan and Einstein frame}\label{appenJorEin}

\begin{figure}[t]
     \begin{center}
\includegraphics[scale=0.42]{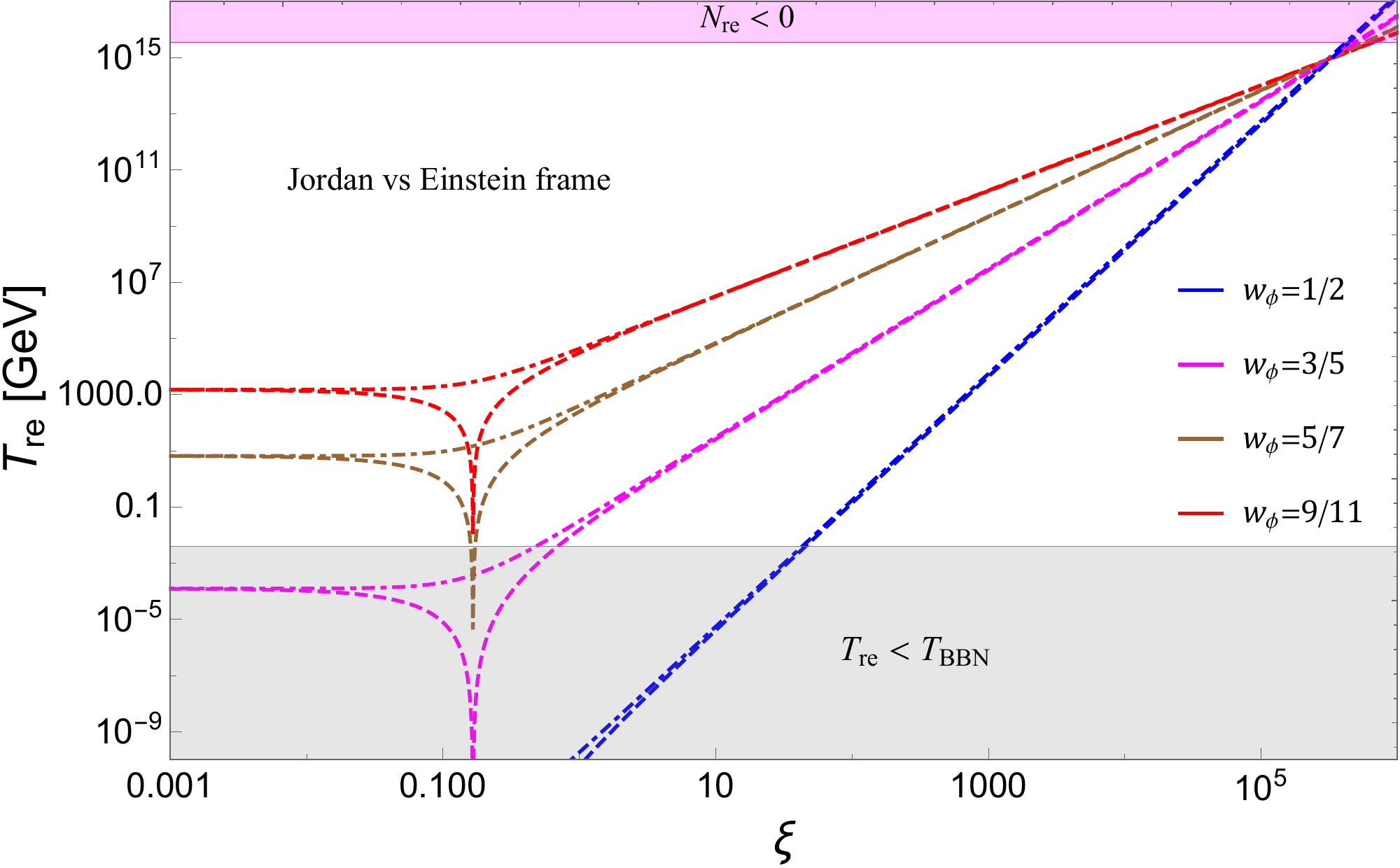}
\caption{\textit{ Figure represents a comparison of $T_{\rm re}$ vs $\xi$ variation for different EoS between the perturbative studies in the Jordan and Einstein frames. 
Dashed lines correspond to the perturbative predictions in the Jordan frame, and the dot-dashed lines correspond to the Einstein frame predictions.
Except near the conformal limit, $\xi\sim 1/6$, it shows almost the same temperature predictions made by the studies in two different frames.}} 
\label{comparisonfigJorEin}
\end{center}
\end{figure}
Based on the production rates given in Eq.(\ref{eq:EJrate}), we have compared the contribution in Jordan and Einstein frames in Fig.(\ref{comparisonfigJorEin}). It clearly shows that in the Einstein frame, the conformal behavior of a massless scalar field is lost and giving rise to a non-vanishing production in the time-dependent background at $\xi = 1/6$. Whereas such production is indeed vanishing in the Jordan frame, as expected.
This indicates a notable departure of the Einstein frame prediction from the Jordan frame and establishes the non-equivalence of these two frames in the context of perturbative reheating predictions. It is interesting to explore such non-equivalence further in the quantum regime.

\bibliographystyle{apsrev4-1}
\bibliography{AYANreferences}

\end{document}